\documentclass[13 pt]{article}
\usepackage{amssymb}

\newcommand{\betah}{\hat{\beta}}

\def\vpbar{\overline \varphi}
\def\zbar{\overline z}
\def\hE{\hat {\cal E}}

\def\vp{\varphi}

\newcommand{\beq}{\begin{equation}}
\newcommand{\eeq}{\end{equation}}
\newcommand{\bea}{\begin{eqnarray*}}
\newcommand{\eea}{\end{eqnarray*}}
\newcommand{\beqa}{\begin{eqnarray}}
\newcommand{\eeqa}{\end{eqnarray}}

\newcommand{\idF}{{id_{\cal F}}}
\newcommand{\idA}{{id_{\cal A}}}

\newcommand{\idAt}{{id_{\tilde{\cal A}}}}

\begin{document}

\newfont{\elevenmib}{cmmib10 scaled\magstep1}%

\newcommand{\Title}[1]{{\baselineskip=26pt \begin{center}
            \Large   \bf #1 \\ \ \\ \end{center}}}
\hspace*{2.13cm}%
\hspace*{1cm}%
\newcommand{\Author}{\begin{center}\large
           Pascal Baseilhac\footnote{
baseilha@phys.univ-tours.fr}
\end{center}}
\newcommand{\Address}{{\baselineskip=18pt \begin{center}
           \it Laboratoire de Math\'ematiques et Physique Th\'eorique CNRS/UMR 6083,\\
Universit\'e de Tours, Parc de Grandmont, 37200 Tours, France
      \end{center}}}
\baselineskip=13pt

\bigskip
\vspace{-1cm}

\Title{Deformed Dolan-Grady relations in quantum integrable
models}\Author

\vspace{- 0.1mm}
 \Address

\vskip 0.6cm

\centerline{\bf Abstract}\vspace{0.3mm} A new hidden symmetry is
exhibited in the reflection equation and related quantum
integrable models. It is generated by a dual pair of operators
$\{\textsf{A}, \textsf{A}^*\}\in{\cal A}$ subject to $q-$deformed
Dolan-Grady relations.  Using the inverse scattering method, a new
family of quantum integrable models is proposed. In the simplest
case, the Hamiltonian is linear in the fundamental generators of
${\cal A}$. For general values of $q$, the corresponding spectral
problem is quasi-exactly solvable. Several examples of
two-dimensional massive/massless (boundary) integrable models are
reconsidered in light of this approach, for which the fundamental
generators of ${\cal A}$ are constructed explicitly and exact
results are obtained. In particular, we exhibit a dynamical
Askey-Wilson symmetry algebra in the (boundary) sine-Gordon model
and show that asymptotic (boundary) states can be expressed in
terms of $q-$orthogonal polynomials. \vspace{1mm}

\vspace{0.1cm}  {\small PACS: 02.20.Uw; 11.30.-j; 11.25.Hf;
11.10.Kk}
\vskip 0.8cm

\vskip -0.6cm

{{\small  {\it \bf Keywords}:  Leonard pair; Tridiagonal pair;
Tridiagonal algebra; Dolan-Grady relations; Onsager algebra;
Quadratic algebras; Massive and massless boundary integrable field
theory}}
%
%
%
%
\section{Introduction}
The exact solution of the planar Ising model in zero magnetic
field \cite{Onsager} has provided a considerable source of
developments in the theory of exactly solvable systems of
statistical mechanics, or quantum field theory in two dimensions.
Onsager's successful approach was originally based on the
so-called Onsager algebra and its representations. Using algebraic
methods, Onsager derived exact results such that the largest and
second largest eigenvalues of the transfer matrix of the model.
Afterwards, the two-dimensional Ising model was reconsidered using
the now famous free fermion (Clifford algebra) techniques
\cite{Kauf}. The Onsager algebra itself being not necessary in
this latter approach, it probably explains why it assumed only a
position of an interesting curiosity in the following years.
Despite of this, in the 1980s the Onsager algebra appeared
\cite{Perk,Davies} to be closely related with the quantum
integrable structure discovered by Dolan and Grady in \cite{DG},
associated with a class of Hamiltonians of the form
\beqa H_{DG} = A_0 + \kappa A_1 \label{H}\eeqa
where $\kappa$ is a coupling constant and $A_0$, $A_1$ are Onsager
algebra fundamental generators. More generally, the Onsager
(in)finite dimensional algebra in the basis of generators
$\{A_m,G_n\}$, $m=0,\pm 1,\pm2,...$ and $n=1,2,...$ reads
\beqa [A_n,A_m]=4G_{m-n}\ ,\qquad [G_m,A_n]=2A_{n+m}-2A_{n-m}\ ,
\qquad [G_m,G_n]=0 \ . \label{Onsager}\eeqa
In \cite{DG}, it was shown that $H_{DG}$ actually belongs to an
(in)finite family of mutually commuting conserved quantities
provided the so-called Dolan-Grady relations
\beqa [A_0,[A_0,[A_0,A_1]]]=16[A_0,A_1]\ ,\qquad
[A_1,[A_1,[A_1,A_0]]]=16[A_1,A_0]\  \label{DG}  \eeqa
are satisfied. As pointed out by Dolan and Grady, the power of the
formulation (\ref{DG}) relies in the operator statement. Indeed,
Hamiltonians of apparently different quantum integrable models can
be written as (\ref{H}) together with (\ref{DG}). For instance, it
is the case for the Ising and XY models \cite{DG}, superintegrable
chiral Potts models \cite{Potts} and some generalizations
\cite{Ahn}.\vspace{1mm}

Although the Onsager algebra (\ref{Onsager}) was the most
important object in \cite{Onsager}, it received less attention in
the following years than the star triangle relations (which did
not play any essential role in original Onsager's work) did.
Actually, the most important progress in the approach of
integrable systems was based on the star-triangle relations which
originated in \cite{Onsager,Wannier} and led to the Yang-Baxter
equations, the theory of quantum groups, as well as the quantum
inverse scattering method. This approach was further extended in
\cite{Skly}: there, it was shown that the Yang-Baxter algebra has
to be enlarged with the reflection equation (RE)  which arises in
various places: factorized scattering theory on the half-line
\cite{Cher}, integrable lattice models with non-periodic boundary
conditions \cite{Skly}, non-commutative differential geometry on
quantum groups. In this generalized framework, the integrability
condition of a model follows from the existence of a solution to
the reflection equation, the so-called reflection matrix. Indeed,
it is the basic object which enters in the definition the quantum
transfer matrix \cite{Skly} denoted $\tau(u)$ below. Using the
fact that, by construction, $\big[ \tau(u_1),\tau(u_2) \big]=0$
for all $u_1$, $u_2$ and expanding in the spectral parameter $u$,
the quantum transfer matrix provides a generating function for an
(in)finite number of integrals of motion in the model.\vspace{1mm}

Whereas the inverse scattering method is well understood and
provides many non-perturbative results in integrable models
(spectrum, scattering amplitudes,...), a formulation based on an
(in)finite dimensional algebra of the Onsager-type (\ref{Onsager})
would be obviously highly desirable: it would provide a new
algebraic approach to integrable {\it massive}
models\,\footnote{The importance of such infinite dimensional
symmetry might be compared to the Virasoro algebra with generators
$L_n$ which play a crucial role in the analysis of conformal field
theory \cite{BPZ}.}. Consequently, an interesting question is
whether such algebraic structure might be actually hidden in the
Yang-Baxter/reflection equations. One reason to believe so, is
that similarly to the relations (\ref{DG}) the explicit
expressions of the $R-$matrix and $K-$matrix rely on the (usually
non-local) quantum symmetry behind a model (lattice, quantum field
theory,...) but not on its explicit microscopic (local) details.
As all known models solved using the Onsager
algebra's approach can also be analyzed in the framework of the
inverse scattering method, it is expected that such a link can be
exhibited.\vspace{1mm}

The purpose of this paper is to make a step towards the explicit
construction of an (in)finite dimensional symmetry in massive
integrable models, along the line initiated in \cite{Onsager,DG}.
Indeed, we will show that known integrable models with an
underlying $U_{q^{1/2}}(\widehat{sl_2})$ quantum group symmetry,
for instance the sine-Gordon quantum field theory or XXZ spin
chain, admit an alternative description\,\footnote{Except if
explicitly specified, throughout the paper we assume generic
values of $q$. Among the special cases, one finds the $N=2$
supersymmetric point in the sine-Gordon model which deserves
special attention (for the case with a boundary, see
\cite{Nepo1,BK2}).} based on a ``Hamiltonian'' of the form
\beqa \textsf{H} = \textsf{A} + \textsf{A}^*\ \label{qH}\eeqa
where $q$ is a deformation parameter and the pair of operators
$\textsf{A},\textsf{A}^*$ (sometimes called a tridiagonal pair
\cite{Terwilli1}) satisfies the ``$q-$deformed'' Dolan-Grady
relations
\beqa
[\textsf{A},[\textsf{A},[\textsf{A},\textsf{A}^*]_q]_{q^{-1}}]=\rho[\textsf{A},\textsf{A}^*]\
,\qquad
[\textsf{A}^*,[\textsf{A}^*,[\textsf{A}^*,\textsf{A}]_q]_{q^{-1}}]=\rho^*[\textsf{A}^*,\textsf{A}]\
\label{qDG} . \eeqa
Here we denote the $q$-commutator$ \
[\textsf{A},\textsf{B}]_q=q^{1/2}\textsf{A}\textsf{B}-q^{-1/2}\textsf{B}\textsf{A}$.
Notice that these relations can be easily obtained from slightly
more general ones called the tridiagonal relations
\cite{Terwilli1} (see also\cite{Ter0,Ter1}). Also, the special
case $q=1$ in (\ref{qDG}) leads to the Dolan-Grady relations
(\ref{DG}) \cite{DG} (see also \cite{Perk,Davies}). As we are
going to see, the ``Hamiltonian'' (\ref{qH}) differs from the
usual {\it local} one, and is expressed in terms of {\it nonlocal}
objects. Furthermore, it is the first of an infinite set of
mutually commuting conserved quantities. On one hand, similarly to
the undeformed case (\ref{DG}) the relations (\ref{qDG}) lead to
an operator statement: they provide the integrability condition
for models of the form (\ref{qH}) which does not depend on the
dimension of the system or the nature of the space-time manifold.
On the other hand, as we shall see the explicit
realization/representation of the operators
$\textsf{A},\textsf{A}^*$ will depend on the model under
consideration.\vspace{2mm}

The paper is organized as follows. In Section 2, we derive a
general solution (so-called $K-$matrix) of the RE in the
spin$-1/2$ representation of $U_{q^{1/2}}(\widehat{sl_2})$. In a
first part, assuming that the non-commuting entries $K_{ij}(u)$ of
this $K-$matrix are Laurent polynomials of degree $-2\leq d\leq 2$
in the spectral parameter $u$ we obtain $K_{ij}(u)\in {\cal
F}un(u;A,A^*)$ where $A,A^*$ is a ``Leonard'' pair subject to
$q-$deformed Dolan-Grady relations (\ref{qDG}). More general
$K-$matrices, denoted $K^{(n)}(u)$ below, are obtained using a
standard ``dressing'' procedure.

In Section 3, in order to derive in dependently the previous
$K-$matrix we study the affine quantum group symmetry of the
reflection equation. It is shown to be associated with a coaction
$\varphi$ acting on Leonard pairs which preserves the algebraic
structure of the reflection equation and generates all
$K^{(n)}(u)$ starting from $K^{(0)}(u)\equiv K(u)$. A set of
relations for a (universal) $K-$matrix are also proposed. In some
sense, they extend the well-known ones of Drinfeld for the
$R-$matrix. For the spin$-\frac{1}{2}$ representation and using
this approach, we explicitly derive $K(u)$. As expected, it agrees
with the one obtained by solving  directly the reflection
equation.

In Section 4, we consider the previous family of solutions
$K^{(n)}(u)$ of the RE in the framework of the generalized inverse
scattering method for the simplest case $n=0$. We obtain naturally
a generalization of the result of Dolan-Grady \cite{DG} for
quantum integrable systems. Indeed, the corresponding Hamiltonian
(\ref{qH}) is a linear combination of $\textsf{A}$ and
$\textsf{A}^*$ satisfying (\ref{qDG}), and enjoys a duality
property exchanging $\textsf{A}\leftrightarrow \textsf{A}^*$. Its
spectral problem is quasi-exactly solvable, except if $q$ is a
root of unity in which case it becomes exactly solvable. We give
two simple explicit examples of representations for
$\textsf{A},\textsf{A}^*$ satisfying (\ref{qDG}). For a single
cyclic representation, the form (\ref{qH}) coincides with the
Azbel-Hofstadter Hamiltonian on a square lattice in the chiral
gauge which spectrum is already known. For a single infinite
dimensional representation, one finds Askey-Wilson polynomials as
eigenfunctions of (\ref{qH}).

In Section 5, we give some explicit examples of two-dimensional
quantum integrable field theory which contain non-local operators
satisfying (\ref{qDG}). In a first part, we point out that some
particular combinations of non-local conserved charges in the
sine-Gordon quantum field theory actually satisfy the relations
(\ref{qDG}), and in some special cases generate the Askey-Wilson
algebra. It follows that an infinite set of non-local conserved
quantities in {\it involution} can be obtained. The spectral
problem for the simplest one (\ref{qH}) and higher conserved
quantities is found to be quasi-exactly solvable and allows us to
relate one-particle states to Askey-Wilson polynomials.

In a second part, we apply more general cases $n\neq 0$ of
$K^{(n)}(u)$ to two-dimensional boundary quantum integrable models
coupled with boundary degrees of freedom located at the boundary,
which {\it must} satisfy (\ref{qDG}) in order to ensure
integrability. Related integrable models are, for instance, the
boundary sine-Gordon model in the massive phase
\cite{GZ,Delius,BK} or massless limit \cite{Saleur,BLZ} coupled
with various realizations of $\textsf{A},\textsf{A}^*$. It follows that the boundary
space of state admits a basis in terms of Askey-Wilson
polynomials. Finally, we briefly discuss the weak-strong coupling
duality phenomena which relates the bulk algebraic structure
(non-local conserved charges) to the boundary degrees of freedom.
We find that both operators satisfy the same algebraic relations.
Concluding remarks follow in the last Section.

\section{Quadratic algebras and deformed Dolan-Grady relations}
Among the known examples of quadratic algebraic structures, one
finds the Yang-Baxter algebra. For further analysis, let us first
recall some known results. This algebra consists of a couple
$R(u)$, $L(u)$ where the $R-$matrix solves the Yang-Baxter
equation (YBE) and the $L$-operator satisfies the quadratic
relations (with spectral parameters $u,v$)
\beqa R_{{\cal V}_0{\cal V}'_0}(u/v){L}_{{\cal V}_0}(u)\otimes
{L}_{{\cal V}'_0}(v)={L}_{{\cal V}'_0}(v)\otimes{L}_{{\cal
V}_0}(u)R_{{\cal V}_0{\cal V}'_0}(u/v)\ \label{RLL}
 \eeqa
where ${\cal V}_0,{\cal V}'_0$ denote finite dimensional auxiliary
space representations. Here, the entries of the $L$-operator act
on a quantum space denoted ${\cal V}$ below. If one considers a
two-dimensional (spin$-\frac{1}{2}$) representation for ${\cal
V}_0$ and ${\cal V}'_0$, a solution of the YBE is provided by (in
general, we will denote $R_{{\cal V}_0{\cal V}_0}(u)\equiv R(u)$)
\beqa R(u)=\sum_{i,j\in \{0,3,\pm\}}\omega_{ij}(u)\
\sigma_i\otimes\sigma_j\ , \label{R} \eeqa
where $\sigma_j$ are Pauli matrices, $\sigma_\pm=(\sigma_1\pm
i\sigma_2)/2$ and $\omega_{ij}(u)$ are some combinations of
elliptic functions. The corresponding $L-$operator in (\ref{RLL})
can be obtained in analogy of (\ref{R}). For inequivalent
auxiliary and quantum spaces, it reads
\beqa L(u)=\sum_{j=\{0,3,\pm\}}\omega_{ij}(u)\ \sigma_i\otimes
S_j\ \label{Ldemi}\eeqa
where the algebraic properties of the operators $\{S_j\}$ are
determined by the Yang-Baxter algebra (\ref{RLL}). Here, the
operators $\{S_j\}$ act on the quantum space ${\cal V}$. Comparing
terms with different dependence  on the spectral parameters $u,v$
in (\ref{RLL}) one arrives at a set of algebraic relations for
$\{S_i\}$, a quadratic algebra which is known as the Sklyanin
algebra. However, a simpler solution of the YBE is obtained from
the trigonometric limit of $\{\omega_{ij}(u)\}$ in which case the
$R-$matrix takes the form (\ref{R}) with
\beqa \omega_{00}(u)&=&\frac{1}{2}(q^{1/2}+1)(u-q^{-1/2}u^{-1})\
,\qquad
\omega_{33}(u)= \frac{1}{2}(q^{1/2}-1)(u+q^{-1/2}u^{-1})\ ,\nonumber \\
\omega_{+-}(u)&=&\omega_{-+}(u)=q^{1/2}-q^{-1/2}\
.\label{coeffR}\eeqa
The corresponding limit of the Sklyanin algebra, also called
trigonometric Sklyanin algebra (TSA), is also quadratic. However,
it can be shown to coincide with the quantum enveloping algebra
$U_{q^{1/2}}(sl_2)$ with the identification
\beqa S_0=\frac{q^{s_3/2}+q^{-s_3/2}}{q^{1/4}+q^{-1/4}}\ ,\qquad
\quad S_3=\frac{q^{s_3/2}-q^{-s_3/2}}{q^{1/4}-q^{-1/4}}\
\nonumber\eeqa
where
\beqa [s_3,S_\pm]=\pm S_\pm \ ,\qquad \mbox{and}\qquad
[S_+,S_-]=\frac{q^{s_3}-q^{-s_3}}{q^{1/2}-q^{-1/2}}\
,\label{uqsl2}\eeqa
together with the Casimir operator
\beqa w = q^{1/2}q^{s_3} + q^{-1/2}q^{-s_3} +
(q^{1/2}-q^{-1/2})^2S_-S_+ \ .\label{Casimirsl2}\eeqa
%

An other example of quadratic algebra is provided by the
reflection equation (sometimes called boundary Yang-Baxter
equation). This equation arises in various context (see for
instance \cite{Cher,Kor,Skly,Frei}). Without spectral parameter, a
systematic study of quadratic algebras associated with the RE can
be found in \cite{Kul1,KulSas}. Here we are interested in the
spectral parameter dependent form which appeared originally in
\cite{Cher}.  For ${\cal V}_0={\cal V}'_0$ it reads
\beqa R(u/v)\ (K(u)\otimes I\!\!I)\ R(uv)\ (I\!\!I \otimes K(v))\
= \ (I\!\!I \otimes K(v))\ R(uv)\ (K(u)\otimes I\!\!I)\ R(u/v)\ .
\label{RE} \eeqa
Similarly to (\ref{Ldemi}), in the spin$-\frac{1}{2}$
representation ${\cal V}_0$ we introduce a $K-$matrix of the form
\beqa K(u) = \sum_{j\in\{0,3,\pm\}} \ \sigma_j \otimes
\Omega_j(u)\ . \label{K}\eeqa
We assume that $\Omega_j(u)$ are Laurent polynomials of degree
$-2\leq d \leq 2$ in the spectral parameter $u$ with coefficients
in a (yet unknown) non-commutative associative algebra ${\cal A}$.
Replacing (\ref{K}) in (\ref{RE}), certain coefficients are
vanishing which leads to consider
\beqa \Omega_0(u)&=&   (F+{\tilde G})u/2 - (G+{\tilde F})u^{-1}/2\
, \qquad \qquad \Omega_+(u)=   U u^2 + V u^{-2} + W \ ,\nonumber \ \\
 \qquad \qquad \qquad
\Omega_3(u)&=& (F-{\tilde G})u/2 - (G-{\tilde F})u^{-1}/2\ ,\qquad
\qquad \Omega_-(u)= {\tilde U} u^2 + {\tilde V} u^{-2} + {\tilde
W} \ \ \label{defdep}\eeqa
where $\{U,V,{\tilde U},{\tilde V},F,G,{\tilde F},{\tilde
G},W,{\tilde W}\}\in {\cal A}$. Replacing in (\ref{RE}) and after
some manipulations we find for instance
\beqa &&\big[U,V\big]=\big[U,W\big]=\big[V,W\big]=\big[{\tilde
U},{\tilde V}\big]=\big[{\tilde U},{\tilde W}\big]=\big[{\tilde
V},{\tilde W}\big]=0\ ,\nonumber \\
&& \big[U,{\tilde U}\big]=\big[V,{\tilde V}\big]=0\
,\nonumber\\
&&\big[U,F\big]=\big[U,G\big]=\big[V,F\big]=\big[V,G\big]=
\big[{\tilde U},F\big]=\big[{\tilde U},G\big]=\big[{\tilde
V},F\big]=\big[{\tilde V},G\big]=0\ ,\nonumber \\
&& U{\tilde V}=V{\tilde U}, \quad {\tilde U}V={\tilde V}U \
,\nonumber\eeqa
and similarly substituting $F\leftrightarrow {\tilde
G},G\leftrightarrow {\tilde F}$. It immediately follows that
$\{U,V,{\tilde U},{\tilde V}\}$ belong to the center of ${\cal
A}$. The fixed value of these elements over any quantum space
representation ${\cal V}$ leads to introduce the linear relations
\beqa {\tilde U}=c_0 U \ , \qquad {\tilde V}=c_0 V \
\nonumber\eeqa
with $c_0\in C\!\!\!\!I$. Then, from (\ref{RE}) we also get
\beqa &&(1-q^{-1})(U{\tilde W}-c_0UW)-[F,G]=0\ ,\qquad
(q-1)(V{\tilde W}-c_0VW)-[F,G]= 0\ , \nonumber \\
&&{\tilde F}U-FV={\tilde F}{\tilde U}-F{\tilde V}={\tilde
G}V-GU={\tilde G}{\tilde V}-G{\tilde U}=0\ ,\nonumber \\
&&\big[F,G\big]-[{\tilde F},{\tilde G}]=0\ .\label{rel1}
 \eeqa
Assuming that ${\tilde W},W$ are linearly independent, these
relations imply
\beqa V=q^{-1}U \ ,\qquad {\tilde F}= q^{-1}F \qquad {\tilde G}= q
G\ .\nonumber\eeqa
Let us now turn to bilinear expressions in $F,G,W,{\tilde W}$
following from (\ref{RE}). Using (\ref{rel1}), they reduce to
\beqa UF(1-q^{-2})+qWG-GW&=&0\ ,\qquad
\ \ UG(1-q^{2})+WF-qFW=0\ ,\nonumber\\
c_0UF(1-q^{-2})+qG{\tilde W}-{\tilde W}G&=&0\ ,\qquad
c_0UG(1-q^{2})+F{\tilde W}-q{\tilde W}F=0\ ,\nonumber\\
\big[W,{\tilde W}\big]+ (q-q^{-1})(q^{-1}F^2-q G^2)&=&0 \
.\label{rel2}\eeqa
The elements $\{W,{\tilde W}\}\in {\cal F}un\big(F,G\big)$ may be
any polynomials in $\{F,G\}$.  However, coefficients of various
terms are subjects to the remaining first equation of (\ref{rel1})
and  eqs. (\ref{rel2}). Consistency implies that the allowed
combinations are reduced to
\beqa W=c_1[F,G]_{q^{-1}}+c_2\ \qquad \mbox{and}\qquad {\tilde
W}=-c_0 c_1[F,G]_{q}+c_0c_2 \ ,\eeqa
where $\{c_1,c_2\}\in C\!\!\!\!I$, together with the value of the
element (in the center of ${\cal A}$) fixed to $U=-1/c_0
c_1q^{-1/2}(q-q^{-1})$. It also follows that $\{F,G\}$ are subject
to the trilinear relations
\beqa &&FG^2+G^2F-(q+q^{-1})GFG -\frac{q^{-1}}{c_0
c_1^2}F=\frac{c_2}{c_1}(q^{-1/2}-q^{1/2})G\
,\nonumber\\
&&GF^2+F^2G-(q+q^{-1})FGF -\frac{q}{c_0
c_1^2}G=\frac{c_2}{c_1}(q^{-1/2}-q^{1/2})F\ .\label{tri} \eeqa
Notice that these relations are sufficient in order to satisfy the
last eq. in (\ref{rel2}). Setting
\beqa F\equiv q^{1/2}{A} \qquad \mbox{and}\qquad G\equiv
q^{-1/2}{A}^* \ \nonumber\eeqa
in (\ref{tri}), it is interesting to observe that such kind of
trilinear algebraic relations are called the Askey-Wilson
relations first introduced by Zhedanov in \cite{Zhed1}. They also
appear in the context of representation theory associated with
tridiagonal algebras (\ref{qDG}) (for more details, see
\cite{Ter2}). According to \cite{Terwilli1} an ordered pair
$A,A^*$ of linear transformations on a finite
dimensional\,\footnote{It is also possible to find infinite
dimensional or cyclic representation ${\cal V}$. Although examples
can be found in \cite{Terwilli1} and below, Leonard pairs in these
cases remain to be define accordingly.} vector space ${\cal V}$
\beqa A\ :\ \  {\cal V}\ \rightarrow\  {\cal V}\ \qquad
\mbox{and}\qquad \ \ A^*\ : \ \ {\cal V}\ \rightarrow\ {\cal
V}\label{LP} \eeqa
is a Leonard pair if there exists a basis for ${\cal V}$ with
respect to which the matrix representing $A$ is diagonal and the
matrix representing $A^*$ is irreducible tridiagonal\,\footnote{A
square matrix $X$ is called tridiagonal whenever each nonzero
entry lies on either the diagonal, subdiagonal or the
superdiagonal. It is irreducible tridiagonal whenever each entry
on the subdiagonal or superdiagonal is nonzero.}, and similarly
exchanging $A \leftrightarrow A^*$. A Leonard pair is essentially
the same thing as a tridiagonal pair for which both operators are
diagonalizable with all eigenspaces of dimension one
[\cite{Terwilli1}, Lemma 2.2]. The Leonard pairs are classified in
[\cite{Ter2}, Theorem 1.9] and [\cite{Ter3}, Theorem 5.16]. In
\cite{Terwilli2}, it is shown that a Leonard pair $A,A^*$
satisfies
\beqa A^*A^2 + {A^2}A^* - (q+q^{-1})AA^*A - \gamma (AA^*+A^*A) -
\rho A^* &=& \gamma^* A^2 + \omega A + \eta I\ \nonumber ,\\
A{A^*}^2 + {A^*}^2A - (q+q^{-1})A^*AA^* - \gamma^* (AA^*+A^*A) -
{\rho^*} A &=& \gamma {A^*}^2 + \omega A^* + \eta^* I
\label{nonstand}\ \eeqa
for a unique sequence of parameters
$\gamma,\gamma^*,\rho,\rho^*,\omega,\eta,\eta^*$. To relate the
quadratic algebra (\ref{RE}) and (\ref{tri}) to the algebraic
relations (\ref{nonstand}) associated with a Leonard pair
${A},{A}^*$, we identify\,\footnote{Note that a Leonard pair, say
${\tilde A},{\tilde A}^*$, satisfying (\ref{nonstand}) can be
deduced from a different pair $A,A^*$ associated with
(\ref{nonstand}) for $\gamma=\gamma^*=\eta=\eta^*=0$ and
(\ref{qDG}) using following substitutions in (\ref{nonstand}): $A
\rightarrow {\tilde A}\equiv A+\alpha I$\ ,$A^* \rightarrow
{\tilde A}^*\equiv A^* +\alpha^*I$\ and $\gamma \rightarrow
-(q^{1/2}-q^{-1/2})^2\alpha$, \ $\rho \rightarrow \rho
+(q^{1/2}-q^{-1/2})^2\alpha^2$,\ $\omega \rightarrow \omega +
2(q^{1/2}-q^{-1/2})^2\alpha\alpha^*$, \ $\eta \rightarrow
-\rho\alpha^* -\omega\alpha
-(q^{1/2}-q^{-1/2})^2\alpha^2\alpha^*$\  where
$\{\alpha,\alpha^*\}\in C\!\!\!\!I$.}
$\gamma=\gamma^*=\eta=\eta^*=0$ and
\beqa \rho=\rho^*\equiv \frac{1}{c_0c_1^2} \  , \quad \omega\equiv
-\frac{c_2}{c_1}(q^{1/2}-q^{-1/2})\ \quad \mbox{and}\qquad
\gamma=\gamma^*=\eta=\eta^*=0\ ,\label{liencoef}\eeqa
where the parameters $c_0,c_1\neq 0$, $c_2$ are arbitrary.
Combining all previous expressions, we conclude that any solution
of the reflection equation (\ref{RE}) of degree $-2\leq d \leq 2$
in the spectral parameter $u$  - with non-commuting entries - can
be written in the form (\ref{K}) where
\beqa \Omega_0(u)&=&   ({A} + {A}^*) (q^{1/2}u-q^{-1/2}u^{-1})/2 \
,\qquad \qquad \Omega_3(u)= ({A} - {A}^*)
(q^{1/2}u+q^{-1/2}u^{-1})/2 \ ,\nonumber \\
\Omega_+(u)&=& \quad   -\frac{q^{1/2} u^2 + q^{-1/2} u^{-2}}{c_0c_1(q-q^{-1})}
 - c_1[{A}^*,{A}]_{q} +c_2\ ,\nonumber \\
\Omega_-(u)&=&-\frac{q^{1/2} u^2 + q^{-1/2} u^{-2}}{c_1(q-q^{-1})}
- c_0c_1[{A},{A}^*]_{q} +c_0c_2 \ .\label{solfin}\eeqa

More general solutions of the reflection equation (\ref{RE}) can
now be obtained from (\ref{K}) with (\ref{solfin}) using the
``dressing'' procedure proposed by Sklyanin in \cite{Skly} that we
apply here: Knowing a solution of the reflection equation
(\ref{RE}), say $K^{(0)}(u)$, with quantum space ${\cal
V}^{(0)}\equiv {\cal V}$ and given the Lax operator
$L_{\verb"j"}(u)$ given by (\ref{Ldemi}) with (\ref{coeffR}),
(\ref{uqsl2}) acting on a quantum space ${\cal V}_{\verb"j"}$ we
define a family of ``dressed'' reflection matrix
\beqa K^{(n)}(u)\equiv L_{\verb"n"}(uk)\cdot\cdot\cdot
L_{\verb"1"}(uk)K^{(0)}(u)L_{\verb"1"}(uk^{-1})\cdot\cdot\cdot
L_{\verb"n"}(uk^{-1})\ \label{Kn}\eeqa
acting on the quantum space
\beqa {\cal V}^{(n)}\equiv
\bigotimes_{{\verb"j"}={\verb"1"}}^{\verb"n"} {\cal V}_{\verb"j"}
\otimes {\cal V}\ .\label{ten}\eeqa
Provided $[S_j,a]=0$ for all $a\in {\cal A}$\ , it is easy to show
that $K^{(n)}(u)$ satisfies the RE (\ref{RE}) for any
value\,\footnote{For $k=1$ and up to an overall spectral parameter
dependent factor one has $L^{-1}(u)\sim L(u^{-1})$.} of the
parameter $k$. Choosing $K^{(0)}\equiv K(u)$ defined by (\ref{K})
with (\ref{solfin}) and (\ref{nonstand}), (\ref{liencoef}), we
finally obtain a whole family of solutions of the RE (\ref{RE}).

To conclude, let us mention that finite/infinite dimensional
irreducible representations of $K^{(n)}(u)$ in the
spin$-\frac{1}{2}$ auxiliary space representation ${\cal V}_0$ are
classified according to the tensor product (\ref{ten}). As the
tensor product of representations ${\cal V}_{\verb"j"}$
(associated with $U_{q^{1/2}}(sl_2)$) is well understood, it is
important to know all the Leonard pairs as well as their cyclic
generalizations. In Section 4 and 5, we will provide explicit
examples of such representations, related with integrable systems.

\section{Quantum affine reflection algebra}
Given any finite dimensional auxiliary space representation ${\cal
V}_0={\cal V}'_0$  (for instance the spin$-\frac{1}{2}$
representation) it is well-known that the (four dimensional)
matrix solution of the Yang-Baxter equation
\beqa R(u/v)R(u)R(v)=R(v)R(u)R(u/v) \label{RRR} \eeqa
can be derived using the underlying symmetry algebra. For the
tensor product of two vector representations, the corresponding
$R-$matrices were derived by Jimbo \cite{Jimbo} for all
non-exceptional quantum affine Lie algebras. In particular, using
the $U_{q^{1/2}}(\widehat{sl_2})$ symmetry in the
spin$-\frac{1}{2}$ representation the $R-$matrix (\ref{R}) with
(\ref{coeffR}) follows.

An analogous method was initiated in \cite{Nepo} and further
developed in \cite{NepoD,Mac,Del} in order to derive some
$K-$matrix solutions of the RE (\ref{RE}) using certain coideal
subalgebras of quantum Kac-Moody algebras or Yangians \cite{Mac}.
There are, however, known examples of $K-$matrix derived using
different symmetries that do not fit in this framework. For
instance, in \cite{BK} we introduced an example of  ``dynamical''
extension of coideal subalgebra of $U_{q^{1/2}}(\widehat{sl_2})$.
The purpose of this section is to reconsider from a more general
point of view
 the meaning of this extension in the context of
$U_{q^{1/2}}(\widehat{sl_2})-$comodules. In particular, we propose
a rather general approach to the construction of a universal
$K-$matrix based on (\ref{univ}).

\subsection{Preliminaries} The universal $R-$matrix associated
with the quantum Kac-Moody algebra $U_{q^{1/2}}(\widehat{sl_2})$
was found by Drinfeld \cite{Drin} using the quantum double
construction. This quantum Kac-Moody algebra, denoted for
simplicity ${\cal F}$ below, is generated by the elements
$\{h_j,e_j,f_j\}$, $j\in \{0,1\}$. They satisfy the commutation
relations
\beqa [h_i,h_j]=0\ , \quad [h_i,e_j]=a_{ij}e_j\ , \quad
[h_i,f_j]=-a_{ij}f_j\ ,\quad
[e_i,f_j]=\delta_{ij}\frac{q^{h_i/2}-q^{-h_i/2}}{q^{1/2}-q^{-1/2}}\
\nonumber\eeqa
together with the $q-$Serre relations
\beqa [e_i,[e_i,[e_i,e_j]_{q}]_{q^{-1}}]=0\ ,\quad \mbox{and}\quad
[f_i,[f_i,[f_i,f_j]_{q}]_{q^{-1}}]=0\ . \eeqa
The sum $k=h_0+h_1$ is the central element of the algebra. The
Hopf algebra structure is ensured by the existence of a
comultiplication $\Delta_{{\cal F}}: {\cal F}\rightarrow {\cal
F}\otimes {\cal F}$ and a counit ${\cal E}_{{\cal F}}: {\cal
F}\rightarrow C\!\!\!\!I$ with
\beqa \Delta_{{\cal F}}(e_i)&=&e_i\otimes q^{h_i/4} +
q^{-h_i/4}\otimes e_i\ \nonumber \\
 \Delta_{{\cal F}}(f_i)&=&f_i\otimes q^{h_i/4} + q^{-h_i/4}\otimes f_i\ ,\nonumber\\
 \Delta_{{\cal F}}(h_i)&=&h_i\otimes I\!\!I + I\!\!I \otimes h_i\ \label{coprod}
\eeqa
and\vspace{-0.3cm}
\beqa {\cal E}_{{\cal F}}(e_i)={\cal E}_{{\cal F}}(f_i)={\cal
E}_{{\cal F}}(h_i)=0\ ,\qquad {\cal E}_{{\cal F}}({I\!\!I})=1\
.\label{counit}\eeqa
Furthermore, ${\cal F}$ is a coalgebra \cite{Chari} because
\beqa (\idF \times {\cal E}_{{\cal F}} )\circ \Delta_{{\cal F}}
\cong \idF\ , \qquad ({\cal E}_{{\cal F}} \times \idF)\cong \idF\
,\qquad (\Delta_{{\cal F}} \times \idF) \circ \Delta_{{\cal F}} =
(\idF \times \Delta_{{\cal F}})\circ \Delta_{{\cal F}} \
.\nonumber \eeqa

Letting $R\in {\cal F}\otimes {\cal F}$ denotes the universal
$R-$matrix, as shown in \cite{Drin} it is the unique invertible
element which satisfies the relations:
\beqa &&\Delta_{{\cal F}}'(x)=R\Delta_{{\cal F}}(x)R^{-1}\quad
\mbox{for all} \quad x\in {\cal F} \ ,\nonumber\\
 &&(\idF\times \Delta_{{\cal
F}})(R)=R_{{\verb"13"}}R_{{\verb"12"}}\ ,\qquad(\Delta_{{\cal F}}
\times \idF )(R)=R_{{\verb"13"}}R_{{\verb"23"}}\
\label{Drinfeld}\eeqa
where $\Delta_{{\cal F}}'=\sigma\circ\Delta_{{\cal F}}$ and one
introduces the permutation map $\sigma(x\otimes y )=y\otimes x$
for all $x,y\in{\cal F}$. Here, using the definition $R=\sum_i
a_i\otimes b^i$  we denote $R_{{\verb"12"}}=\sum_i a_i\otimes
b^i\otimes I\!\!I$, $R_{{\verb"13"}}=\sum_i a_i\otimes I\!\!I
\otimes b^i$ and $R_{{\verb"23"}}=\sum_i I\!\!I \otimes a_i\otimes
b^i$. In order to obtain a solution of the Yang-Baxter equation
(non-linear in the entries of $R$ for some finite dimensional
representation), it is sufficient to find $R\in {\cal F}\otimes
{\cal F}$ that satisfies (\ref{Drinfeld}). Indeed, from the first
equation of (\ref{Drinfeld}) one writes
\beqa R_{{\verb"12"}}(\Delta_{{\cal F}}\times \idF)(x\otimes
y)=(\Delta_{{\cal F}}'\times\idF)(x\otimes y)R_{{\verb"12"}}\qquad
\mbox{for all} \quad x,y\in {\cal F} \label{RDrin} \ .\eeqa
Setting $R\equiv x\otimes y$ and using the last relation of
(\ref{Drinfeld}) one obtains the Yang-Baxter equation\vspace{1mm}
\beqa
R_{{\verb"12"}}R_{{\verb"13"}}R_{{\verb"23"}}=R_{{\verb"23"}}R_{{\verb"13"}}R_{{\verb"12"}}\
.\label{univYBE}\eeqa
The universal version of (\ref{RLL}) is derived similarly. One
denotes the element $L\in {\cal F} \otimes {\tilde {\cal A}}$,
where $\tilde {\cal A}$ is a quadratic associative algebra.
Actually, suppose that this quadratic algebra is such that there
exists an element $L\equiv x\otimes {\tilde a}$ which satisfies
(\ref{RDrin}) substituting $\idF\rightarrow \idAt$. Then $L$
satisfies the Yang-Baxter algebra\vspace{1mm}
\beqa
R_{{\verb"12"}}L_{{\verb"1"}}L_{{\verb"2"}}=L_{{\verb"2"}}L_{{\verb"1"}}R_{{\verb"12"}}\
.\label{univYBA}\eeqa
Also, assuming a Hopf algebra structure for ${\tilde {\cal A}}$
with comultiplication $\Delta'_{\tilde {\cal A}}: {\tilde {\cal
A}} \rightarrow {\tilde {\cal A}} \otimes {\tilde {\cal A}}$, more
general solutions $L^{(n)}$ of (\ref{univYBA}) can be obtained. To
construct them, one introduces the $n-$comultiplication
\beqa \Delta'^{(n)}_{\tilde {\cal A}}: \ {\tilde {\cal A}}
\longrightarrow {\tilde {\cal A}} \otimes \cdot\cdot\cdot \otimes
{\tilde {\cal A}}\ \quad , \qquad \Delta'^{(n)}_{\tilde {\cal
A}}\equiv (id_{\tilde {\cal A}}\times \cdot\cdot\cdot \times
id_{\tilde {\cal A}} \times \Delta'_{\tilde {\cal A}})\circ
\Delta'^{(n-1)}_{\tilde {\cal A}}\
 \eeqa
for $n\geq 3$ with $\Delta'^{(2)}_{\tilde {\cal A}}\equiv
\Delta'_{\tilde {\cal A}}$, and define
\beqa L^{(n)}\equiv(id_{{\cal F}}\times \Delta'^{(n)}_{\tilde
{\cal A}})(L)=L_{\verb"n"}\cdot\cdot\cdot L_{\verb"1"} \label{Ln}\
\eeqa
where indices here refer to the quantum space ${\tilde {\cal A}}$.
Acting with $(id_{{\cal F}}\times \Delta'^{(n)}_{\tilde {\cal
A}})$ on (\ref{univYBA}) and using the standard comultiplication
rules, it is straightforward to show that $L^{(n)}$ is indeed a
solution of (\ref{univYBA}). Obviously, we may have started from
$\Delta_{\tilde {\cal A}}$ instead. In that case, one shows that
${\overline L}^{(n)}\equiv L_{\verb"1"}\cdot\cdot\cdot
L_{\verb"n"}$ also solves (\ref{univYBA}). In other words, the
comultiplication $\Delta'_{\tilde {\cal A}}$ (or similarly
$\Delta_{\tilde {\cal A}}$) preserves the algebraic structure
associated with (\ref{univYBA}).\vspace{2mm}

Later on, we will focus on the spin$-\frac{1}{2}$ representation
of $U_{q^{1/2}}(\widehat{sl_2})$ and central element $k=0$.
Denoting $\pi^{(\frac{1}{2})}_u(x)$ the representation  with
spectral parameter $u$ of the element $x\in {\cal F}$, one has
\beqa (\pi^{(\frac{1}{2})}_u \times
\pi^{(\frac{1}{2})}_v)[R_{\verb"12"}]=R(u/v)\ , \qquad
(\pi^{(\frac{1}{2})}_u \times \idAt)[L]=L(u) \label{repR12}\eeqa
and more generally
\beqa (\pi^{(\frac{1}{2})}_u \times \idAt \times \cdot\cdot\cdot
\times \idAt)[L^{(n)}]&=&L_{\verb"n"}(u)\cdot\cdot\cdot
L_{\verb"1"}(u)\ \label{repLn}.
\eeqa
In particular, the representations (\ref{repR12}) are used in
(\ref{univYBE}) and (\ref{univYBA}) to obtain the Yang-Baxter
equation (\ref{RRR}) or the Yang-Baxter algebra (\ref{RLL}),
respectively. In the following subsection, we will proceed by
analogy for the reflection equation.

\subsection{Comodule algebra and the reflection equation}
In this part, we study the algebraic structure related with the
spectral parameter dependent reflection equation (\ref{RE}). For
$F_q(GL(2))$ i.e. without spectral parameter, part of the
following analysis was considered by Kulish and Sklyanin in
\cite{Kul1} (see also \cite{KulSas}, \cite{Gervais} and
\cite{Alg}). First, let us recall the concept of coaction and
comodule necessary in the following analysis. Then, we propose a
quite general framework in order to derive a ``universal''
reflection matrix. The special case of spin$-\frac{1}{2}$
representation is treated in full details.\vspace{3mm}

{\bf Definition:} \cite{Chari} {\it Given a coalgebra
${\tilde{\cal F}}$ with comultiplication $\Delta_{{\tilde{\cal
F}}}$ and counit ${\cal E}_{{\tilde{\cal F}}}$, ${\cal A}$ is
called a left ${\tilde{\cal F}}-$comodule if there exists a left
coaction map $\zeta:\ \ {\cal A}\rightarrow {\tilde{\cal F}}
\otimes {\cal A}$ such that}
\beqa (\Delta_{{\tilde{\cal F}}} \times \idA)\circ
\zeta=(id_{{\tilde{\cal F}}}\times\zeta)\circ\zeta\ ,\qquad
({\cal E}_{{\tilde{\cal F}}} \times \idA)\circ \zeta \cong \idA\
.\label{defcoaction}\eeqa
Right ${\tilde{\cal F}}-$comodules are defined similarly.\vspace{1mm}

By analogy with the
previous subsection, we are now looking for coaction maps
$\varphi$ (and $\varphi'$) which preserve the
algebraic structure associated with the quadratic algebra
(\ref{RE}). Similarly to (\ref{Ln}), these coaction maps must be
such that $K^{(n)}$ is generated from $K^{(0)}$. For a given (here
spin$-\frac{1}{2}$) representation of ${\cal F}$, we previously
obtained the solutions $K^{(n)}(u)$ in terms of $K^{(0)}(u)$. For
instance, setting $n=1$ in (\ref{Kn}) and replacing $k\rightarrow
v$, it leads us to define the coaction maps by
\beqa (\pi^{(\frac{1}{2})}_u \times \pi_v \times \idA)[(\idF\times
\varphi)(K)]&\equiv& L(uv) K(u) L(uv^{-1})\ ,\nonumber\\
(\pi^{(\frac{1}{2})}_u \times \pi_v \times \idA)[(\idF\times
\varphi')(K)]&\equiv& L(uv^{-1}) K(u) L(uv)\ .\label{LKL}\eeqa
Here the notation $\pi_v$ means that we do not specify the
dimension of the representation for ${\tilde{\cal A}}$. From now
on, we identify ${\tilde{\cal A}}\equiv{\cal F}$. Then, it is
important to remember that the $L-$operators can be written in
terms of the universal $R-$matrix as
\beqa L(uv^{-1})=(\pi^{(\frac{1}{2})}_u \times
\pi_v)[R_{\verb"12"}] \qquad \mbox{and}\qquad
L(uv)=(\pi^{(\frac{1}{2})}_u \times \pi_v)[R_{\verb"21"}]\
\label{represL}\eeqa
in order to introduce relations analogous\,\footnote{However, a
rigorous proof of the existence of a unique universal reflection
matrix remains to be done.} to (\ref{Drinfeld}):\vspace{5mm}

{\bf Conjecture:} {\it Let $\Delta_{\cal F}$ be the
comultiplication associated with the coalgebra ${\cal F}$. Let
$\varphi$ (and $\varphi'$) be a two-sided coaction associated with an associative
algebra ${\cal A}$. There exists an element $K\in {\cal F} \otimes
{\cal A} $ which satisfies the relations:}
\beqa (i) &&\varphi'(a)=K\varphi(a)K^{-1}\quad \mbox{for all}
\quad  a\in {\cal A} ,\nonumber \\
(ii)&&(\Delta_{\cal F}\times \idA)(K)=(K\otimes
I\!\!I)R_{{\verb"21"}}(I\!\!I\otimes K)\
,\qquad (\Delta'_{\cal F}\times \idA)(K)=(I\!\!I \otimes K)R_{{\verb"21"}}(K \otimes I\!\!I )\ ,\nonumber\\
(iii) &&(\idF\times \varphi)(K)=R_{\verb"21"}(K\otimes
I\!\!I)R_{{\verb"12"}}\ ,\ \ \qquad \qquad( \idF \times
 \varphi')(K)=R_{{\verb"12"}}(K\otimes I\!\!I )R_{{\verb"21"}}\
 .\label{univ}\eeqa\vspace{-3mm}\\
Notice that the relation $(ii)$ already appeared in
\cite{Gervais}. If such an element $K$ exists, it is then
straightforward to show that this element satisfies the reflection
equation. There are two possible ways. For instance, from $(i)$ we
write
\beqa (\idF\times \varphi')(x\otimes a)(I\!\!I \otimes K)=(I\!\!I
\otimes K)(\idF\times \varphi)(x\otimes a)\quad \mbox{for all}
\quad x\otimes a \in {\cal F}\otimes {\cal A}\
,\label{met1}\nonumber\eeqa
or similarly from the first relation of (\ref{Drinfeld}) one has
\beqa R_{{\verb"12"}}(\Delta'_{\cal F}\times \idA)(x\otimes
a)=(\Delta_{\cal F}\times \idA)(x\otimes a)R_{{\verb"12"}}\quad
\mbox{for all} \quad x\otimes a \in {\cal F}\otimes {\cal A}\
.\label{met2}\nonumber\eeqa
Choosing $K\equiv x\otimes a$ and using $(iii)$ or $(ii)$,
respectively, in the two equations above we deduce the universal
form of the reflection equation
\beqa R_{{\verb"12"}}(K\otimes I\!\!I)R_{{\verb"21"}}(I\!\!I
\otimes K)=(I\!\!I \otimes K)R_{{\verb"21"}}(K \otimes I\!\!I
)R_{{\verb"12"}}\ .\label{univRE}\eeqa
Similarly to the previous subsection, a
generalized $n-$coaction $\varphi^{(n)}$ can be introduced
for $n\geq 2$ with $\varphi^{(1)}\equiv\varphi$. Together with (\ref{LKL}), it
can be used to generate the family of solutions (\ref{Kn})
starting from $K(u)=K^{(0)}(u)$. Consequently, the basic solution
(\ref{K}) with (\ref{solfin}) is a fundamental object.\vspace{3mm}

Here, we will not try to construct the universal reflection
matrix. Instead, we will consider in details the derivation of
$K(u)$ in (\ref{K}) in the spin$-\frac{1}{2}$ representation of
$U_{q^{1/2}}(\widehat{sl_2})$, using (\ref{univ}). Note that a
special choice of representation for the associative algebra
${\cal A}$ will not be necessary. The spectral parameter dependent
form of $R$, $K$ and $L$ in the spin$-\frac{1}{2}$ representation
is given by (\ref{repR12}) and
\beqa (\pi^{(\frac{1}{2})}_u \times
\pi^{(\frac{1}{2})}_v)[R_{\verb"21"}]=R(uv)\ \quad \mbox{and}\quad
(\pi^{(\frac{1}{2})}_u \times \idA)[K]=K(u)\label{repK}\ \eeqa
from which (\ref{RE}) follows. For higher dimensional
representation one gets obviously the same form.

Now, we would like to find the explicit expression of the action
of $\varphi$ on the elements of ${\cal A}$. To do that, let us
expand (\ref{LKL})  in
terms of the spectral parameter $v$.  There, we use the explicit
form of $K^{(0)}(u)$ given by (\ref{K}) together with
(\ref{Ldemi}). After some straightforward calculations, we find
that the $U_{q^{1/2}}({\widehat{sl_2}})$ algebra arises explicitly
through the evaluation homomorphism
$U_{q^{1/2}}(\widehat{sl_2})\longrightarrow U_{q^{1/2}}({sl_2})$
\beqa &&\pi_v[e_1]= vS_+\ , \qquad \ \ \ \pi_v[e_0]= vS_-\ , \nonumber\\
&&\pi_v[f_1]=
v^{-1}S_-\ ,\qquad \pi_v[f_0]= v^{-1}S_+\ ,\nonumber\\
\ &&\pi_v[q^{h_1/2}]= q^{s_3}\ ,\qquad \ \pi_v[q^{h_0/2}]=
q^{-s_3}\ .\nonumber\eeqa
The expansion of (\ref{LKL}) finally gives  
\beqa (\pi^{(\frac{1}{2})}_u \times \pi_v \times \idA)[(\idF\times
\varphi)(K)]_{11}&=& (\pi_{vq^{1/4}} \times \idA)[ qu^3\zeta({A})-q^{-1}u^{-3}\zeta({A}^*) +
\zeta'(u,u^{-1};{A},{A}^*)]\ ,\nonumber\\
(\pi^{(\frac{1}{2})}_u \times \pi_v \times \idA)[(\idF\times
\varphi')(K)]_{22}&=& (\pi_{vq^{1/4}} \times \idA)[ qu^3\zeta({A^*})-q^{-1}u^{-3}\zeta({A}) +
\zeta'(u,u^{-1};{A}^*,{A})]\
,\nonumber\\
(\pi^{(\frac{1}{2})}_u \times \pi_v \times \idA)[(\idF\times
\varphi)(K)]_{12}&=& (\pi_{vq^{1/4}} \times \idA)[ -\frac{qu^4 + q^{-1}u^{-4}}{c_0c_1
(q-q^{-1})} \zeta(I)+
\zeta''(u^2,u^{-2};{A},{A}^*)]\ ,\nonumber\\
(\pi^{(\frac{1}{2})}_u \times \pi_v \times \idA)[(\idF\times
\varphi)(K)]_{21}&=& (\pi_{vq^{1/4}} \times \idA)[ -\frac{qu^4 + q^{-1}u^{-4}}{c_1 (q-q^{-1})}
\zeta(I)+ {\overline{\zeta}}''(u^2,u^{-2};{A}^*,{A})]
 \label{expandon}\eeqa
where\vspace{-3mm}
\beqa &&\zeta(I)=I\!\!I\otimes I\ ,\nonumber\\
&&\zeta({A})\ = (c_+e_1 + c_-f_1)q^{h_1/4}\otimes I +
q^{h_1/2}\otimes {A}\ ,\nonumber \\
&&\zeta({A}^*)= (c_-e_0 + c_+f_0)q^{h_0/4}\otimes I +
q^{h_0/2}\otimes {A}^*\ .\label{defphi}\eeqa
The terms $\zeta'$, $\zeta''$, ${\overline{\zeta}}''$ do not
contribute in the asymptotic $u\rightarrow\infty$ or $u\rightarrow
0$ \ so we don't write them explicitly here. The parameters $c_+$,
$c_-$ read
\beqa c_+=-1/c_0c_1(q^{1/2}+q^{-1/2})\ \qquad \mbox{and}\qquad
c_-=-1/c_1(q^{1/2}+q^{-1/2})\ .\label{cpm}\eeqa
It should be stressed that each of these maps
$\zeta,\zeta',\zeta''$ define a left coaction and confirm the fact
that ${\cal A}$ is a left $U_{q^{1/2}}(\widehat{sl_2})-$comodule.
For instance, it is easy to check using (\ref{coprod}),
(\ref{counit}) that $\zeta$ defined by (\ref{defphi}) is
consistent with the comultiplication $\Delta'_{\cal F}=\sigma\circ
\Delta_{\cal F }$, i.e. satisfies (\ref{defcoaction}). Starting
from $(i)$ in (\ref{univ}), we can now write (the representation
in $v-$space is not specified yet)
\beqa (\pi^{(\frac{1}{2})}_u \times \pi_v \times \idA)[(I\!\!I
\otimes K)(\idF\times \varphi)(x\otimes a)] =
(\pi^{(\frac{1}{2})}_u \times \pi_v \times \idA)[(\idF\times
\varphi')(x\otimes a)(I\!\!I \otimes K)]\label{eqr}\eeqa
for all $x\in {\cal F}$, $a\in {\cal A}$. Setting $K=x \otimes a$
and using $(iii)$ of (\ref{univ}) together with (\ref{represL}),
(\ref{LKL}), both sides can be expanded in the spectral parameter
$u$. In the spin$-\frac{1}{2}$ representation, setting
$u\rightarrow \infty$ or $u\rightarrow 0$ it follows that the
leading terms are given in terms of (\ref{defphi}). We find that
the $K-$matrix (\ref{K}) must satisfy
\beqa
K(v)\pi^{(\frac{1}{2})}_{vq^{1/4}}[\zeta(a)]=\pi^{(\frac{1}{2})}_{v^{-1}q^{1/4}}[\zeta(a)]K(v)
\ \qquad \mbox{for all} \qquad a\in\{I,{A},{A}^*\}\ .\label{inter}
\eeqa
This intertwining equation is actually {\it sufficient} to
determine $K(u)$. Indeed, in this representation one has $\pi^{(\frac{1}{2})}(S_\pm)=\sigma_\pm$\ ,
$\pi^{(\frac{1}{2})}(s_3)=\sigma_3/2$.
From the diagonal part of (\ref{inter}), it is easy to get the
expressions $\Omega_0(v)$ and $\Omega_3(v)$ in (\ref{solfin}).
Replacing them in the off-diagonal part of (\ref{inter}), it
follows that $K_{ij}(v)\in {\cal F}un(v;A,A^*)$ for $i\neq j$  are
polynomials of degree $-2,2,0$ in the spectral parameter $v$.
Also, one finds easily that the spectral parameter dependent terms
must commute with $A,A^*$, whereas the remaining terms (denoted
$W,{\tilde W}$ in (\ref{defdep})) must satisfy for instance:
\beqa q^{1/2}A^* {\tilde W} - q^{-1/2}{\tilde W}A^* &=&
-c_-A(q^{1/2}+q^{-1/2})\ ,\nonumber \\
q^{-1/2}A {\tilde W} - q^{1/2}{\tilde W}A &=&
c_-A^*(q^{1/2}+q^{-1/2})\ \label{Weq} \eeqa
and similarly for ${W}$ using $c_+$. However, from the diagonal
part of (\ref{inter}), one also finds $c_+({\tilde
W}-c_0{W})=[{A},{A}^*]$. Replacing in (\ref{Weq}) one immediately
gets the trilinear algebraic relations (\ref{nonstand}) together
with (\ref{liencoef}). Consequently, the
 solution in the spin$-\frac{1}{2}$ representation (\ref{K}) with
(\ref{solfin}), (\ref{nonstand}), (\ref{liencoef}) can be derived
directly from $(i)$ in (\ref{univ}) as soon as one knows the
explicit form of the coaction acting on ${\cal A}$, or at least
its asymptotic behavior in the ``auxiliary'' spectral parameter.
To resume, the coaction plays here a role analogous to the
coproduct in case of a Hopf algebra structure. 

\section{General features of the integrable structure}
Either considering the constraints which follow from the
reflection equation, or studying the quantum affine reflection
symmetry, we have seen that the trilinear algebraic relations
(\ref{nonstand}) with (\ref{liencoef}) arise in both approaches.
However, it is clear that the parameters $c_0,c_1,c_2$ entering in
the reflection matrix (\ref{K}) are not restricted by any symmetry
argument during this process. Instead, their explicit values will
depend on the explicit realization of the Leonard pair $A,A^*$
i.e. on {\it model-dependent} and {\it representation}
characteristics of the corresponding operators. These
realizations, as we will see below, admit finite/infinite
dimensional or cyclic irreducible representations. Consequently,
to speak in full generality i.e without specifying a
realization/representation one should instead of (\ref{nonstand})
consider the $q-$deformed Dolan-Grady relations (\ref{qDG}) as the
integrability condition for the class of models associated with
(\ref{RE}).\vspace{1mm}

Having identified this integrability condition, the next step is
the construction of related integrable models. In the quantum
inverse scattering framework, there are many examples of quantum
integrable systems which are constructed starting solely from a
$L-$operator solution of the Yang-Baxter algebra (\ref{RLL}).
Using the coproduct structure, it follows that the entries of
(\ref{Ln}) (see for instance in the Appendix) are rather nonlocal
objects, usually expressed in terms of the local degrees of
freedom of the system acting on the quantum space. Taking the
trace over the auxiliary space of the monodromy matrix, one
obtains among conserved quantities the Hamiltonian of the
system. By comparison between $L$ and $K$ i.e. (\ref{RLL}) and
(\ref{RE}), we can then consider a quantum integrable model
constructed solely from a $K-$operator of the form (\ref{K}) with
(\ref{solfin}) i.e. choosing $n=0$ in (\ref{Kn}). Its generating
function is given by
\beqa \tau(u)\equiv Tr_{{\cal V}_0}\big[K(u)\big]\label{tau0}\ .
\eeqa
Explicitly one finds \ $\tau(u) = (uq^{1/2}-u^{-1}q^{-1/2})
{\textsf{H}}$ \ where the Hamiltonian\,\footnote{Obviously, using
a scale transformation of Leonard pairs one can introduce extra
parameters related with $\rho,\rho^*$ in (\ref{qH}).} takes the
form (\ref{qH}) with $\textsf{A}\rightarrow A$,
$\textsf{A}^*\rightarrow A^*$ which satisfy the $q-$deformed
Dolan-Grady relations (\ref{qDG}). Beyond the Hamitonian, note
that there exists a Casimir operator
\beqa Q=q\rho^*{A}^2 +
q^{-1}\rho{A^*}^2-\frac{(q-q^{-1})^2}{4}[A,A^*]_q[A,A^*]_{q^{-1}}
+\frac{(q-q^{-1})}{2}\omega(AA^*+A^*A)\ . \eeqa

An interesting problem is now to construct a family of quantum
integrable models associated with (\ref{qH}) in the spirit of
\cite{DG} based on more general objects, say
$\textsf{A},\textsf{A}^*$, satisfying (\ref{qDG}) but {\it not}
(\ref{nonstand}). First, by analogy with the undeformed case
\cite{Davies} it is indeed well expected that one
can find operators $\textsf{A},\textsf{A}^*$ which admit two types of representations:\\
(I)\ \ - ``Single'' representations i.e. $\textsf{A},\textsf{A}^*$
are represented by a pair $A,A^*$ which obey (\ref{nonstand});\\
(II)\ - ``Multiple'' representations (non-trivial tensor products
of irreducible ones) i.e. $\textsf{A},\textsf{A}^*$ satisfy
(\ref{qDG}), {\it not} (\ref{nonstand}).\\
Note that representations of type (II) can be obtained from more general solutions (\ref{Kn}) of the reflection equation, as shown 
in \cite{Tridiag}. Secondly,
the power of the algebraic constraint (\ref{qDG}) relies on the
fact that it is an operator statement which does not refer to any
local structure of the model under consideration. It is thus
natural to expect that $\textsf{A},\textsf{A}^*$ are rather
nonlocal objects. By analogy with the undeformed case \cite{DG}, it is possible to
show that there exists an (in)
finite family of mutually commuting
conserved quantities of the form ($r\geq 0$)
\beqa G^{(2r)}(\textsf{A},\textsf{A}^*)= \textsf{A}^{(2r)} +
\textsf{A}^{*{(2r)}}\ \qquad \mbox{with}\qquad
G^{(0)}(\textsf{A},\textsf{A}^*)\equiv \textsf{H} \label{G2r}\eeqa
provided (\ref{qDG}) is satisfied. For example, assuming
(\ref{qDG}) it is not difficult to check that
\beqa
G^{(2)}(\textsf{A},\textsf{A}^*)=\big[\textsf{A},\big[\textsf{A},\textsf{A}^*\big]_{q^{-1}}\big]_q
+
\big[\textsf{A}^*,\big[\textsf{A}^*,\textsf{A}\big]_{q^{-1}}\big]_q\
\label{G2} \eeqa
is conserved. We report the detailed construction of higher
conserved quantities to a separate work \cite{Tridiag}. Then, an important
problem concerns the diagonalization of these conserved quantities
$G^{(2r)}(\textsf{A},\textsf{A}^*)$. It is actually closely
related with the problem of partial algebraization of the spectrum
associated with difference equations which arises in the context
of quasi-exactly solvable systems. Two situations
arise:\vspace{1mm}

(1) $q$ is  root of unity: the spectrum associated with
$G^{(2r)}(\textsf{A},\textsf{A}^*)$ is finite. All periodic
eigenfuntions can be obtained algebraically;\vspace{1mm}

(2) $q$ is not a root of unity: the spectral problem associated
with $G^{(2r)}(\textsf{A},\textsf{A}^*)$ is quasi-exactly
solvable. It is related with the representation theory of
(\ref{qDG}). Only a part of the spectrum can be obtained, with
polynomial eigenfunctions (for instance Askey-Wilson, big $q$-Jacobi
polynomials,...) or their restrictions
\cite{Terwilli1}.\vspace{2mm}

Using the explicit realization of the operators
$\textsf{A},\textsf{A}^{*}$ in terms of known
algebras (see below), for single representations (type (I) above)
the spectral problem associated with (\ref{G2r}) for $r\geq 0$ can be reduced to
a single $q-$difference equation of the form
\beqa a(z)\Psi(qz) + d(z)\Psi(q^{-1}z) -v(z)\Psi(z) =
\Lambda\Psi(z)\ \label{spectral}\eeqa
where $a(z)$, $d(z)$, $v(z)$ are polynomials in $z$ defined
according to the model under consideration, the
realization/representation of the Leonard pair and the explicit
form of $G^{(0)}$. The eigenfunctions $\Psi(z)$ of the
algebraized part of the spectrum are also polynomials of finite
order, Askey-Wilson polynomials for instance. Then, $\Lambda$
denotes the corresponding subset of eigenvalues of the operator
$G^{(0)}(\textsf{A},\textsf{A}^*)$.\vspace{1mm}

To be more explicit, let us give two examples corresponding to the
cases (1) and (2) above and single representations (type (I)). For
instance, following our results in \cite{BK} it is not difficult
to find a realization for the operators $\textsf{A},\textsf{A}^*$
in terms of the Weyl algebra. Choosing
\beqa \textsf{A}= -i\frac{\rho^{1/2}}{q-q^{-1}}(Q+Q^{-1})\qquad
\mbox{and} \qquad \textsf{A}^* =
-i\frac{{\rho^*}^{1/2}}{q-q^{-1}}(P+P^{-1}) \qquad \mbox{where}
\qquad  PQ= q^{-1} QP \ ,\label{Weyl} \eeqa
one can check that $\textsf{A},\textsf{A}^*$ satisfy (\ref{qDG})
and {\it also} (\ref{nonstand}) for
$\gamma=\gamma^*=\omega=\eta=\eta^*=0$. For $q=\exp(2i\pi M/N)$
root of unity where $M,N$ are mutually prime integers, it is
well-known that the Weyl algebra can be realized using $N\times N$
matrices with the matrix elements $\langle
m|Q|n\rangle=q^{-m}\delta_{m,n}$ and $\langle
m|P|n\rangle=\delta_{m+1,n\ (mod\ N)}$ for which one has
$P^N=Q^N=1$. In this case, the pair $\textsf{A},\textsf{A}^*$
admits a finite dimensional (cyclic) representation but the matrix
representing $\textsf{A}^*$ is not tridiagonal. For these values
of $q$, the Hamiltonian (\ref{qH}) together with (\ref{Weyl})
actually coincides with the Azbel-Hofstadter (AH) Hamiltonian
\cite{Hofs} which describes the problem of Bloch electrons in a
magnetic field on a two-dimensional lattice. This model has been
studied in details in several works\,\footnote{For an alternative
derivation of the AH Hamitonian starting from a $L-$operator
instead of $K$, we report the reader to the Appendix.} (see for
instance \cite{Wieg,Fad,Flor,Nic}). Using the relation between the
group of magnetic translations and the quantum group $U_q(sl_2)$,
it was shown that the AH Hamiltonian can be written as a linear
combination of the quantum group generators \cite{Wieg}. Also, its
spectrum can be represented in terms of solutions of Bethe ansatz
equations on high genus algebraic curves \cite{Wieg,Fad}. In
particular, for $q$ a root of unity the polynomial eigenfunctions
and the Bethe ansatz cover all the spectrum. Previous analysis
shows that the AH Hamiltonian on a square lattice in the chiral
gauge can be written as a linear combination of operators of the
form (\ref{Weyl}), the operators $P,Q$ being identified with the
generators of magnetic translations in each direction of the
two-dimensional lattice. Using the bilinear formulation of
(\ref{nonstand}) in terms of the Askey-Wilson algebra
\cite{Zhed4}, it also explains the relation between the zero mode
wave function and AW polynomials pointed out in \cite{Wieg}.
Consequently, the AH Hamiltonian provides an example of quantum
integrable system associated with the $q-$deformed Dolan-Grady
relations (\ref{qDG}).\vspace{2mm}

Our second example is quasi-exactly solvable i.e. corresponds to
generic values of $q$. For simplicity, let us consider the values
$\rho=-(q-q^{-1})^2$ and $\rho^*=-abcdq^{-1}(q-q^{-1})^  2$ in
(\ref{qDG}). Leonard pairs in this case admit an infinite
dimensional representation in the basis of the Askey-Wilson
polynomials with continuous weight\,\footnote{For finite
dimensional representation, it leads to Askey-Wilson polynomials
with discrete weights.} \cite{Terwilli1}. These $q-$orthogonal
polynomials, here denoted $\textsc{P}_{n}(z)$ with integer $n\geq
0$, are symmetric Laurent polynomials of the variable $z$ i.e.
they are invariant under the change $z\rightarrow z^{-1}$. They
can be written explicitly in terms of the basic $q-$hypergeometric
functions as\vspace{2mm}
\beqa \textsc{P}_{n}(z)=\ _4\Phi_3 \left (
\begin{array}{c}
  q^{-n},\ abcdq^{n-1}, \ az, \ az^{-1}\\
  ab, \ ac, \  ad \\
\end{array}; q,q
\right)\ .\label{polyAW}\eeqa
In particular, these polynomials are known to satisfy the three
terms recurrence relation
\beqa \beta_n\textsc{P}_{n+1}(z) + \alpha_n\textsc{P}_{n}(z)
+\gamma_n\textsc{P}_{n-1}(z)
=(z+z^{-1})\textsc{P}_{n}(z)\label{diseq}\eeqa
where $\textsc{P}_{-1}\equiv 0$, $\textsc{P}_{1}\equiv 1$ and
\beqa \beta_n
&=&\frac{(1-abq^n)(1-acq^n)(1-adq^n)(1-abcdq^{n-1})}{a(1-abcdq^{2n-1})(1-abcdq^{2n})}\
,\nonumber \\
\gamma_n
&=&\frac{a(1-q^n)(1-bcq^{n-1})(1-bdq^{n-1})(1-cdq^{n-1})}{(1-abcdq^{2n-2})(1-abcdq^{2n-1})}\
,\nonumber \\
\alpha_n &=& a+a^{-1}-\beta_n -\gamma_n \
.\label{coefpoly}\nonumber\eeqa
In this basis and choosing the symmetric case $abcd=q$, one has
\beqa \pi(\textsf{A})=\left(
        \begin{array}{ccccc}
\alpha_0 & \gamma_1 & &  \\
\beta_0 & \alpha_1 & \gamma_2 & \\
& \beta_1 & \alpha_2 & \cdot & \\
& & \cdot & \cdot & \cdot  \\
&  & & \cdot & \cdot
\end{array}\right)\ \qquad \pi(\textsf{A}^*)=\mbox{diag}\left(
2, q+q^{-1}, q^{2}+q^{-2}, \cdot,\cdot,\cdot\right) \
.\label{repmat} \eeqa

On the other hand, these polynomials are eigenvectors
\cite{AWpoly} of a second order $q-$difference operator $I\!\!D$
with parameters $a,b,c,d$:
$I\!\!D\textsc{P}_{n}(z)=(abcdq^{n-1}+q^{-n})\textsc{P}_{n}(z)$. The Leonard pair in this representation becomes
$\pi(\textsf{A})=z+z^{-1}$, $\textsf{A}^*=I\!\!D$. Setting
$abcd=q$, it follows that the Askey-Wilson polynomials
(\ref{polyAW}) satisfy the $q-$difference equation
\beqa \xi(z)\textsc{P}_{n}(qz) +
\xi(z^{-1})\textsc{P}_{n}(q^{-1}z) =
\big(\xi(z)+\xi(z^{-1})-(1-q^{-n})(1-q^{n})\big)\textsc{P}_{n}(z)\
,\label{diffeq}\eeqa
where
\beqa \xi(z)\equiv
\frac{(1-az)(1-bz)(1-cz)(1-dz)}{(1-z^2)(1-qz^2)}\ .\nonumber\eeqa
Using (\ref{diseq}), (\ref{repmat}) and (\ref{diffeq}), the
spectral problem associated with the Hamiltonian (\ref{qH}) for
$\rho=\rho^*=-(q-q^{-1})2$ takes the form (\ref{spectral})
together with
\beqa a(z)=\xi(z)\ ,\quad d(z)=\xi(z^{-1}) \ \quad \mbox{and}\quad
v(z)= \xi(z)+\xi(z^{-1})-z-z^{-1}-2\ .\eeqa
Using the standard Bethe ansatz technique, it follows that the
roots $z^{(n)}_m$ of the Askey-Wilson polynomial (\ref{polyAW}) -
eigenfunctions of (\ref{qH}) - determine the spectrum and obey the
Bethe ansatz equations. This problem was treated in details in
\cite{Wieg2} so we don't need to repeat the analysis here and
report the reader to this work for details.\vspace{1mm}

Finally, it should be stressed that according to the model and
realization of the Leonard pair considered, the values of the
structure constants $\rho,\rho^*,\eta,\eta^*,\omega$ in
(\ref{nonstand}) determine the form of polynomial eigenfunctions.
Obviously, the two examples above do not exhaust all
possibilities. For instance, if $\omega\neq 0$ and
$\rho=\rho^*=0$, the eigenfunctions reduce to $q-$Hahn polynomials
associated with $_3\Phi_2$. For $\omega=\rho=\rho^*=0$ they
correspond to $q-$Krawtchouk polynomials associated with
$_2\Phi_1$.

\section{Related two-dimensional quantum integrable models}
\subsection{The sine-Gordon model revisited}
For the coupling constant $\betah^2<2$, the bulk sine-Gordon model
in 1+1 Minkowski space-time is massive and integrable. The
particle spectrum consists of a soliton/antisoliton pair
$(\psi_+(\theta),\psi_-(\theta))$ with mass $M$ and neutral
particles, called ``breathers'', $B_n(\theta)$\ \
$n=1,2,...,<\lambda$. As usual, $E=M\cosh\theta$ and
$P=M\sinh\theta$ the on-shell energy and momentum, respectively,
of the soliton/antisoliton. In this model, ${\cal H}$ is the Fock
space of multi-particle states. A general $N$-particles state is
generated by the ``particle creation operators''
$A_{a_i}(\theta_i)$
\beqa |A_{a_1}(\theta_1)...A_{a_N}(\theta_N)\rangle=
A_{a_1}(\theta_1)...A_{a_N}(\theta_N)|0\rangle\label{asymst}\eeqa
where $a_i$ characterizes the type of particle. The commutation
relations between these operators are determined by the $S$-matrix
elements, which describe the factorized scattering theory
\cite{Zam79}. Integrability imposes strong constraints on the
system which imply that the general $S$-matrix factorizes in
two-particle/two-particle amplitudes which satisfy the quantum
Yang-Baxter equations. To determine this basic $S$-matrix, an
alternative approach was proposed in \cite{Ber91} based on the
existence of non-local conserved charges in quantum integrable
field theory. For the sine-Gordon model, the symmetry algebra is
known to be identified with $U_{q_{0}}({\widehat{sl_2}})$ where
the relation between the deformation parameter and the coupling
constant is $q_{0}=\exp(-2i\pi/\betah^2)$. To exhibit this
underlying quantum symmetry, the authors of  \cite{Ber91} used the
Lagrangian representation of the sine-Gordon model:
\beqa {\cal A}_{SG}=\int d^2x
\Big(\frac{1}{8\pi}(\partial_\nu\phi)^2-2\mu\cos(\betah\phi)\Big)
\label{actionSG}\ .\eeqa
Here $\mu$ is a mass scale and $\phi$ is the sine-Gordon field. In
the perturbed conformal field theory (CFT) approach, the
corresponding free bosonic fundamental field is decomposed in its
holomorphic/antiholomorphic components\,\footnote{If we denote the
expectation value over the Fock space vacuum of the CFT by
$\langle ... \rangle$, then these components are normalized such
that $ \langle{\varphi}(z){{\varphi}}(w)\rangle=-\ln(z-w),\quad
\langle{{\bar\varphi}}({\bar z}){{\bar\varphi}}({\bar
w})\rangle=-\ln({\bar z}-{\bar w}),\quad
\langle{\varphi}(z){{\bar\varphi}}({\bar w})\rangle=0$\ .}
\bea \phi(x,y)=\vp(z)+\vpbar(\zbar)\ .\label{dec}\eea
The non-local conserved currents $J_\pm(x,t),{\overline
J}_\pm(x,t)$ of the form
\bea &&
J_{\pm}(x,t)=:\exp({\pm\frac{2i}{\betah}{\varphi}}(z)):\qquad
\mbox{and}\qquad {\bar J}_{\pm}(x,t)=:\exp({\mp
\frac{2i}{\betah}{\vpbar}}(\zbar)):\ \eea
generate the non-local conserved charges denoted
${Q}_{\pm},{\overline Q}_{\pm}$ below. Due to the non-locality of
the currents, at equal time $t$ they possess non-trivial braiding
betweem themselves. Then, together with the ``bulk'' topological
charge
\beqa {\cal T}=\frac{\betah}{2\pi}\int^{\infty}_{-\infty}dx\
\partial_x {\phi} \eeqa
they generate the quantum enveloping algebra $U_{q_0}(\widehat{
sl_2})$ \cite{Ber91}.  Contrary to the local integrals of motion,
these conserved charges are not in involution. Their action over
asymptotic soliton states can be deduced from a field theoretic
approach \cite{Ber91}: for a single asymptotic
state\,\footnote{For the fundamental soliton/antisoliton with
topological charge $\pm 1$, $j=1/2$. Higher values of $j$ occur
for boundstates.} with topological charge $2m$
\beqa |v;m\rangle\qquad \qquad \mbox{where}\qquad \quad
v=\exp(\theta)\ ,\qquad -j\leq m \leq j \label{singrep}\eeqa
in the spin$-j$ representation and fixed rapidity $\theta$ it
reads
\beqa \pi_{v}[{Q}_{\pm}]=ce^{\lambda\theta}S_\pm q_{0}^{\pm s_3}\
,\qquad \pi_{v}[{\overline Q}_{\pm}]=ce^{-\lambda\theta}S_\pm
q_{0}^{\mp s_3} \qquad\mbox{and} \qquad \pi_{v}[{\cal T}]=2s_3\
\label{repcharges} \eeqa
where
\beqa \lambda=\frac{2}{\betah^2}-1 \qquad \mbox{and} \qquad
c^2=-i\frac{\mu}{\lambda^2}(q_{0}^{-2}-1)\ . \nonumber\eeqa
More generally, the action of the non-local conserved charges
${Q}_{\pm},{\overline Q}_{\pm}$ and ${\cal T}$ over a
multi-particle state is obtained using the coproduct rules of
$U_{q_0}(\widehat{ sl_2})$ (see \cite{Ber91} for details). In the
following, we will associate one-particle states to single
representations (type (I)) and multi-particle states to
representations of type (II).

\subsubsection{Askey-Wilson dynamical symmetry and duality}
A natural question is wether certain combinations of non-local
conserved charges might satisfy the relations (\ref{qDG}) and
generate conserved quantities in {\it involution}. In order to
exhibit such non-trivial combinations in the sine-Gordon model, we
use the following trick. According to the analysis of Section 2
and (\ref{Kn}), the simplest (up to an overall factor) non-trivial
``dynamical'' solution of the RE (\ref{RE}) is of the form
(\ref{K}) with (\ref{solfin}). An easy way to find single
representations (type (I)) of operators satisfying (\ref{qDG}) is
then to start from a non-dynamical solution $K^{(0)}(u)$, even
very simple, and to apply (\ref{Kn}) for $n=1$ using the Lax
operator (\ref{Ldemi}). For instance, let us choose the simple
off-diagonal constant solution of (\ref{RE}) given by $K\equiv
(c_+\sigma_++c_- \sigma_-)/(q^{1/2}-q^{-1/2})$ and set
$v\rightarrow v_0q^{-1/4}$ in (\ref{LKL}). It is straightforward
to obtain the solution (\ref{K}) with (\ref{solfin}) through the
identification
\beqa A\equiv c_+ v_0 S_+ q^{s_3/2} + c_- v_0^{-1} S_- q^{s_3/2}
\qquad \mbox{and}\qquad A^*\equiv c_- v_0 S_- q^{-s_3/2} + c_+
v_0^{-1} S_+ q^{-s_3/2}\ .\label{Nomura}\eeqa
Here we use (\ref{cpm}), (\ref{nonstand}) with (\ref{liencoef})
and we find the relation
\beqa c_2(v_0;j)=
\frac{(v_0^2q^{-1/2}+v_0^{-2}q^{1/2})w_j}{c_0c_1(q^{1/2}+q^{-1/2})(q-q^{-1})}\label{c2}\eeqa
where $w_j$ is the Casimir operator (\ref{Casimirsl2}) eigenvalue
\beqa w_j=q^{j+1/2}+q^{-j-1/2}\ .\eeqa

An other useful realization of operators  satisfying (\ref{qDG})
can be obtained from the non-diagonal non-dynamical solution of
the reflection equation\,\footnote{For a similar calculation, see
\cite{Zab}. Note that this non-dynamical solution can also be
obtained from (\ref{solfin}) for $A$, $A^*$ constants.}\cite{DV},
\cite{GZ}. It yields to
\beqa A\equiv c_+ v_0 S_+ q^{s_3/2} + c_- v_0^{-1} S_- q^{s_3/2} +
\epsilon_+q^{s_3}\qquad \mbox{and}\qquad A^*\equiv c_- v_0 S_-
q^{-s_3/2} + c_+ v_0^{-1} S_+ q^{-s_3/2}+
\epsilon_-q^{-s_3}\label{Nomura2}\eeqa
where $\epsilon_\pm$ are free parameters. Actually, it is not
difficult to check that the above simple realizations of $A,A^*$
satisfy (\ref{nonstand}) with $\gamma=\gamma^*=0$, $\rho=\rho^*=
1/c_0c_1^2$ \ and the structure constants
\beqa
 \omega(v_0;j)&=&
-\frac{c_2(v_0;j)+\epsilon_+\epsilon_-c_1(q^{1/2}-q^{-1/2})}{c_1}(q^{1/2}-q^{-1/2})\
,\nonumber \\
\eta(v_0;j)&=&\frac{\epsilon_+(v_0^2q^{-1/2}+v_0^{-2}q^{1/2})-\epsilon_-w_j}{c_0c_1^2(q^{1/2}+q^{-1/2})}\
, \nonumber\\
\eta^*(v_0;j)&=&\eta(v_0;j)|_{\epsilon_+\leftrightarrow
\epsilon_-}\ .\label{omeg} \eeqa
Here, $c_0$, $c_1$ are non-vanishing parameters, $c_2$ is the same
as above (\ref{c2}) and $w_j$ is the Casimir eigenvalue
(\ref{Casimirsl2}). It is now straightforward to identify the
operators satisfying (\ref{qDG}) in the sine-Gordon model. First,
setting
\beqa v_0=\exp(\lambda\theta)\ ,\qquad q^2_{0}= q\ ,\qquad
c_+c_0=c_-=-\frac{1}{c_1(q^{1/2}+q^{-1/2})}=  c\ \nonumber\eeqa
and using (\ref{repcharges}) one can interpret (\ref{Nomura2}) as
combinations of non-local charges acting on a single (one-particle
state) representation (\ref{singrep}) i.e. of type (I). In
particular, the fact that (\ref{Nomura2}) satisfy the trilinear
algebraic relations (\ref{nonstand}) with (\ref{omeg}) implies
(\ref{qDG}). More generally, in order to find
$\textsf{A},\textsf{A}^*$ which do not reduce to a Leonard pair,
one has to consider the action of the non-local conserved charges
over multi-particle states (type (II)). In this case, we propose
to consider the pair
\beqa {\hat Q}_{+}&=&\frac{1}{c_0}Q_{+}+{\bar
Q}_{-}+\epsilon_{+}q_{0}^{{\cal T}}\ \qquad \mbox{and}\qquad {\hat
Q}_{-}=Q_{-}+\frac{1}{c_0}{\bar Q}_{+}+\epsilon_{-}q_{0}^{-{\cal
T}}\ . \label{nonlocSG}\eeqa
To show that $q-$deformed Dolan-Grady relations are satisfied for
{\it any} $N-$particles state representation of (\ref{nonlocSG})
it is necessary and sufficient to show that the coproduct of
$U_{q_0}(\widehat{ sl_2})$ leaves the relations (\ref{qDG})
unchanged. Thus, we have checked explicitly that
 the non-local conserved charges satisfy the $q-$deformed Dolan-Grady relations
 (\ref{qDG}) with the substitution
\beqa \textsf{A}\rightarrow {\hat Q}_{+}\qquad\quad
\mbox{and}\quad\qquad \textsf{A}^*\rightarrow {\hat
Q}_{-}\label{cor}\eeqa
for any $c_0\neq 0$ and free parameters $\epsilon_\pm$ (for details, see \cite{Tridiag}). In this
case, the value of $\rho$, $\rho^*$ in (\ref{qDG}) is
\beqa \rho=\rho^*=\frac{(1+q)(1+q^{-1})}{c_0}\ . \eeqa

We would like to stress that the trilinear algebraic relations
(\ref{nonstand}) only arise in case of a single representation
(one-particle state). This algebra actually admits a bilinear
formulation in terms of the Askey-Wilson algebra AW$(3)$
introduced and studied by Zhedanov in \cite{Zhed1}. It is
generated by the elements $\{K_0,K_1,K_2\}$ which satisfy
\beqa [K_0,K_1]_q=K_2\ ,\qquad \big[K_2,K_0\big]_q=B K_0 + C_1 K_1
+ D_1 I\ , \qquad \big[K_1,K_2\big]_q=B K_1 + C_0 K_0 + D_0 I\
\label{AW3} \nonumber\eeqa
where $B,C_0,C_1,D_0,D_1$ are the structure constants. It is an
exercise to relate these structure constants to the ones in
(\ref{omeg}). According to the analysis of \cite{Zhed1}, one can
obtain irreducible finite dimensional representations depending on
the value of the rapidity and the dimension of the asymptotic
state representation $2j+1$. Following \cite{Zhed1}, it is
possible to find a basis $|\psi_m\rangle$, $-j\leq m\leq j$, for
the asymptotic particle state of rapidity $\theta$ which
diagonalizes $\pi_v\big[{\hat Q}_{+}\big]$. On the other hand, one
can find a basis $|\varphi_s\rangle$, $-j\leq s\leq j$ for the
asymptotic particle state which instead diagonalizes
$\pi_v\big[{\hat Q}_{-}\big]$. Consequently, the conserved
quantities (\ref{nonlocSG}) enjoy a remarkable duality property
which reminds the one pointed out by Dolan-Grady in their paper
\cite{DG}. Since both representations have the same dimension,
both basis are related by a linear transformation
$|\varphi_s\rangle=\sum\langle s|p\rangle|\psi_p\rangle$. As shown
in \cite{Zhed1}, the corresponding overlap functions $\langle
s|p\rangle$ can be explicitly expressed in terms of the
Askey-Wilson polynomials on a finite interval of the real axis.

\subsubsection{Related spectral problem and asymptotic one-particle states}
It is now possible to construct an infinite set of mutually
commuting conserved quantities of the form (\ref{G2r}) in the
sine-Gordon model (for a detailed analysis of a slightly more general integrable structure, see \cite{Tridiag}), the first ones being
\beqa \textsf{H}&=&{\hat Q}_{+}+{\hat
Q}_{-}\ ,\nonumber\\
G^{(2)}&=&\big[{\hat Q}_{+},\big[{\hat Q}_{+},{\hat
Q}_{-}\big]_{q^{-1}}\big]_q + \big[{\hat Q}_{-},\big[{\hat
Q}_{-},{\hat Q}_{+}\big]_{q^{-1}}\big]_q \ .\eeqa
In general, the corresponding spectral problem is rather
complicated. However, acting on a one-particle  state all
expressions drastically simplify. The corresponding spectral
problem for a state of rapidity $\theta$ and topological charge
$2m$ in a representation of dimension $2j+1$ reads for $r=0,1,$...
\beqa G^{(2r)}|v;m\rangle = \Lambda_{2r}(v;j,m)|v;m\rangle\qquad
\mbox{for all}\qquad -j \leq m \leq j\ .\label{specr}\eeqa
Up to an overall $(v,j)-$dependent coefficient which changes
according to the value of $r$, this spectral problem always
reduces to the one for $\textsf{H}$. Indeed, using the trilinear
algebraic relations (\ref{nonstand}) it is not difficult to check
that if $|v;m\rangle$ is an eigenstate of $\textsf{H}$, so it will
be for all higher conserved quantities. For general values of
$q_{0}$, it is a quasi-exactly solvable problem. To study it, we
follow the analysis of \cite{Wieg2}. Expressed in the weight
basis, the generators of $U_{q_{0}}(sl_2)$ leave invariant the
linear space (of $2j+1$ dimension) of polynomials $F(z)$ of degree
$2j$. The lowest/highest weights are identified with $F_0=1$ and
$F_{2j}=z^{2j}$, respectively. One has
\beqa q_{0}^{\pm s_3} F(z)&=&q_{0}^{\mp j}F(q_{0}^{\pm 1}z)\ ,\nonumber\\
S_+
F(z)&=&-\frac{z}{(q_{0}-q^{-1}_{0})}\big(q_{0}^{-2j}F(q_{0}z)-q_{0}^{2j}F(q_{0}^{-1}z)\big)\
,\nonumber\\
S_-
F(z)&=&\frac{1}{z(q_{0}-q_{0}^{-1})}\big(F(q_{0}z)-F(q_{0}^{-1}z)\big)\
. \label{jrep}\nonumber \eeqa
In this basis of polynomials, for general values\,\footnote{For
some special values of the parameters, the degree of the
polynomial eigenfuntions can be different. Here we do not consider
this possibility.} of the parameters we may represent a
one-particle  state with rapidity $\theta$ (recall that
$v=\exp(\theta)$) by
\beqa \langle z|v;m\rangle = \Psi_m(z)\ \qquad \mbox{with}\qquad
\Psi_m(z)=\prod_{n=1}^{2j}\big(z-z^{(m)}_n\big)\
\label{statem}\eeqa
where $z^{(m)}_n$ denote the roots of the $m-$th polynomial. From
(\ref{specr}) and using the representation above, it is
straightforward to obtain a $q-$difference equation of the form
(\ref{spectral}) with the substitution $q\rightarrow q2_0$ where
the coefficients are explicitly given by
\beqa
a(z)&=&\big(-c_+vzq_{0}^{-3j}+c_-v^{-1}z^{-1}q_{0}^{-j}+\epsilon_+(q_{0}-q_{0}^{-1})q_{0}^{-2j}\big)/(q_{0}-q_{0}^{-1})\
,\nonumber \\
d(z)&=&\big(-c_-vz^{-1}q_{0}^{j}+c_+v^{-1}zq_{0}^{3j}+\epsilon_-(q_{0}-q_{0}^{-1})q_{0}^{2j}\big)/(q_{0}-q_{0}^{-1})\
, \nonumber
\\
v(z)&=&\big(-c_+vzq_{0}^{j}+c_-v^{-1}z^{-1}q_{0}^{-j}+c_+v^{-1}zq_{0}^{-j}-c_-vz^{-1}q_{0}^{j}\big)/(q_{0}-q_{0}^{-1})\
.\label{coj}
 \eeqa
The corresponding eigenvalue $\Lambda_0(v;j,m)$ can be deduced
from the results of \cite{Wieg2}. It is expressed in terms of the
roots $z^{(m)}_n$ in (\ref{statem}) which have to satisfy a system
of $2j$ algebraic (Bethe) equations. This system has exactly
$2j+1$ solutions corresponding to $2j+1$ eigenfunctions
$\Psi_m(z)$, $-j\leq m\leq j$. Note that these roots depend on
value of the rapidity $\theta$ and $\epsilon_\pm$.\vspace{2mm}

To resume, the sine-Gordon model possesses an underlying dynamical
symmetry associated with (\ref{qDG}) with (\ref{cor}). For single
representations, it is isomorphic to the quadratic Askey-Wilson
algebra, which is generated by the charges. This symmetry is
preserved at all orders in perturbed CFT framework because the
fundamental non-local charges ${Q}_{\pm},{\overline Q}_{\pm}$ and
${\cal T}$ are conserved. The point is that these charges generate
an infinite number of mutually commuting conserved quantities
$G^{(2r)}$ \cite{Tridiag}. Actually, their actions on asymptotic states mix
non-trivially the holomorphic and antiholomorphic sectors (in the
perturbed CFT approach). Then, conservation of $G^{(2r)}$ for any
$r$ imposes a severe restriction on the asymptotic states
structure. For instance, a one-particle  state (\ref{statem})
admits a representation in terms of polynomials whose roots
satisfy Bethe equations. The corresponding eigenvalues can be
computed using the usual Bethe ansatz techniques (see for instance
\cite{Wieg2}).

\subsection{The sine-Gordon model with/without a dynamical boundary}
Following \cite{Skly}, boundary integrable models on the lattice -
for instance the XXZ spin chain - can be obtained using
(\ref{Kn}). In \cite{Delius} we have applied Sklyanin's formalism
\cite{Skly} to construct a classical integrable field theory
restricted to the half-line, namely the sine-Gordon model, coupled
with a mechanical system at the boundary. In general, such
two-dimensional classical field theory will remain integrable
provided there exists a classical solution $K_{cl}(u)$ of the
classical reflection equation. In this case, the generating
function reads
\beqa \tau_{cl}(u)\equiv Tr_{{\cal
V}_0}\big[T_{cl}(u)K_{cl}(u){\hat
T}_{cl}(u^{-1})\big]\label{taucl}\eeqa
where $T_{cl}(u)$ (and similarly ${\hat T}_{cl}(u)$) is the
sine-Gordon classical monodromy matrix. Expanding in the spectral
parameter $u$ one obtains an infinite number of mutually
``Poisson'' commuting quantities, which give the integrals of
motion and ensure the integrability of the classical system with a
boundary. However, it is clear that the analysis of \cite{Delius}
for the classical boundary sine-Gordon model can be extended to
more general boundary operators provided they satisfy some
classical quadratic algebra.
This can be done starting from the classical limit of the
reflection matrix (\ref{K}).
%
%
Expanding $\tau_{cl}(u)$ in the spectral parameter $u$, among the
integrals of motion one identifies the Hamiltonian. Its boundary
contribution is found to be linear in the classical boundary
operators coupled with the sine-Gordon classical field $\phi$.
Setting the coupling constant of the sine-Gordon model $\betah=0$,
this boundary term reduces to the classical analogue of
(\ref{qH}).\vspace{1mm}

At quantum level, an integrable (massive) boundary sine-Gordon
model coupled with a quantum mechanical system at the boundary was
proposed in \cite{BK}. There, the asymptotic behavior of the
boundary operators was expressed in terms of the operators
(\ref{Weyl}). However, as we have shown in Sections 2 and 3, the
existence of a $K-$matrix is related with (\ref{nonstand}) (see
also \cite{BK} for details). It is thus more general to consider
in full generality boundary operators which asymptotically satisfy
(\ref{nonstand}) rather than choosing a (somehow simpler)
realization like (\ref{Weyl}). Then, we propose
\beqa {\cal A}_{bsg}=\int_{-\infty}^{\infty}dy\int_{-\infty}^{0}dx
\Big(\frac{1}{8\pi}(\partial_\nu\phi)^2-2\mu\cos(\betah\phi)\Big)
+ \mu^{1/2}\int_{-\infty}^{\infty}dy\ \Phi_{pert}^{B}(y)\ + \
{\cal A}_{boundary} \label{action} \eeqa
where the interaction between the sine-Gordon field and the
boundary quantum operators reads
\bea \Phi_{pert}^{B}(y) \ =\ {\cal E}_-(y) e^{i\betah\phi(0,y)/2}
+ {\cal E}_+(y) e^{-i\betah\phi(0,y)/2}\  \eea
and ${\cal A}_{boundary}$ is the kinetic part\,\footnote{A
solution of the ``dual'' reflection (associated with a boundary on
the left hand side) can be used to introduce such term, without
breaking integrability.} associated with the boundary operators
$\hE_\pm(y)$. In the deep UV, this model can be considered as a
relevant perturbation of a conformal field theory (for instance
the free Gaussian field) on the semi-infinite plane with dynamical
boundary conditions at $x=0$. In this case, similarly to the bulk
(\ref{dec}) the free Gaussian field restricted to the half-line
can be also written in terms of its holomorphic/anti-holomorphic
components. However, these components are instead normalized such
that\vspace{-1mm}
\bea \langle{\varphi}(z){{\varphi}}(w)\rangle_0=-2\ln(z-w),\quad
\langle{{\bar\varphi}}({\bar z}){{\bar\varphi}}({\bar
w})\rangle_0=-2\ln({\bar z}-{\bar w}),\quad
\langle{\varphi}(z){{\bar\varphi}}({\bar
w})\rangle_0=-2\ln(z-{\bar w})\  \eea
where $\langle ... \rangle_0$ now denotes the expectation value in
the boundary conformal field theory (BCFT) with Neumann boundary
conditions. The boundary conditions in (\ref{action}) can be
derived from the expectation value of the local field
$\partial_x\phi(0,y)$ with any other local field in first order of
boundary conformal perturbation theory $\mu\rightarrow 0$.
Actually, these quantum boundary conditions take a form similar to
the classical ones \cite{BK}.

We can choose the $y$-direction to be the ``time'' in which case
the Hamiltonian contains the boundary contribution, and the
Hilbert space ${\cal H}_B$ is identified with the half-line
$y=Const.$. Then, correlation functions are calculated over the
boundary ground state denoted $|0\rangle_B$ below. This state can
be expanded as
\beqa |0\rangle_B\ \ =\ \ |vac\rangle_B\otimes|0\rangle_{BCFT}\ \
\ +\ \ \ {\cal O}(\mu) \ ,\label{vac}\eeqa
where \ \ $|0\rangle_{BCFT}\in {\cal H}_{BCFT}$\ belongs to the
Hilbert space of the BCFT and $|vac\rangle_B$ is an eigenstate of
the boundary Hamiltonian for the coupling $\betah\rightarrow 0$.

There are good reason to believe \cite{GZ,BK} that a quantum
analogue of the classical integrals of motion can be constructed
explicitly. Similarly to the bulk case, instead one can show that
the model (\ref{action}) remains integrable at quantum level {\it
too} using the existence of {\it boundary} non-local conserved
charges. Along the line of \cite{Nepo,Mac,BK} it is possible to
construct non-local conserved charges in (\ref{action}) in terms
of the bulk ones (which are no longer conserved individually).
Assuming the same asymptotic behavior at the end point of the time
axis
\beqa {\hat{\cal E}}_{+}(y=\pm\infty)\sim{\hat{\cal E}}_{+}\qquad
\qquad \mbox{and}\qquad \qquad {\hat{\cal
E}}_{-}(y=\pm\infty)\sim{\hat{\cal E}}_{-}\ ,\label{asympt}\eeqa
it follows that the quantum affine reflection symmetry of the
model (\ref{action}) is generated by the operators
\beqa {\hat Q}^{(0)}_{\pm}=Q_{\pm}+{\bar Q}_{\mp}+{\hat{\cal
E}}_{\pm}q_{0}^{\mp {\cal T}_b}\ \label{nonloc}\qquad
\qquad\mbox{with}\qquad \qquad {\cal
T}_b=\frac{\betah}{2\pi}\int^{0}_{-\infty}dx\ \partial_x
{\phi}\eeqa
and \ ${\hat{\cal E}}_{\pm}(y)=\mu^{1/2}{\cal
E}_\pm(y)\betah^2/(1-\betah^2)$\ . For the special case
(\ref{Weyl}), this was shown in details in \cite{BK}. Here,\
${\cal T}_b$ denotes the boundary topological charge which is no
longer independently conserved. In the following, we are going to
consider different restrictions of the boundary operators. This
will reveal interesting features of the model (\ref{action}). In
particular, most of the known and well-studied massive or massless
boundary integrable models can be seen as limiting cases of
(\ref{action}). \vspace{2mm}

$\bullet$ {\bf The boundary sine-Gordon model revisited.}
The quantum boundary sine-Gordon model introduced by
Ghoshal-Zamolodchikov is obtained for the one-dimensional
(trivial) representation of the boundary operators (which become
free parameters) in (\ref{action}). In other words, the boundary
degrees of freedom are ``frozen'' to constants. The corresponding
non-local conserved charges calculated in \cite{Nepo,Mac} take the
form (\ref{nonloc}) together with the substitution
$\hE_\pm\rightarrow \epsilon_\pm$, where $\epsilon_\pm$ are free
parameters\,\footnote{These parameters are expressed in terms of
the boundary parameters in the Lagrangian and some bulk
contributions. I thank Z. Bajnok who attracted my attention to
this point.}. It is then interesting to notice that, from an
algebraic point of view, they coincide exactly with the
spin$-(j=\frac{1}{2})$ representation of (\ref{nonlocSG}) together
with
\beqa c_0=1\ ,\quad {\cal T}_b \rightarrow -{\cal T}\
.\nonumber\eeqa
This should not be surprising as the symmetry between the
holomorphic/antiholomorphic sector must be restored at the
boundary. Due to the presence of the boundary which breaks the
$U_{q_0}({\widehat{ sl_2}} )$ symmetry, the quantum group symmetry
of the boundary sine-Gordon model with fixed boundary conditions
is generated by the non-local conserved charges (\ref{nonlocSG})
for the special case (symmetric) $c_0=1$. In particular, single
representations are associated with a restricted Askey-Wilson
quadratic algebra. As before, the non-local boundary conserved
charges satisfy the $q-$deformed Dolan-Grady relations (\ref{qDG})
with (\ref{omeg}) and $c_0=1$. Conservation of the quantities
$G^{(2r)}$ for $c_0=1$ implies that the boundary space
representation has dimension $2j+1$. A convenient basis for the
boundary (degenerate) ground state $|0\rangle_{B}$ is provided by
the Askey-Wilson polynomials with discrete weights, or can be
written in the same manner than (\ref{statem}). \vspace{1mm}

The boundary scattering properties are known to be encoded in the
boundary reflection matrix, which can be obtained using the
quantum reflection symmetry \cite{Nepo} (see also \cite{Mac}). In
the formalism of Section 3, it corresponds to the choice of
representation $\pi^{(1/2)}\times \pi^{(0)}$ in (\ref{defphi})
where $\pi^{(0)}(A)\sim\epsilon_+$, $\pi^{(0)}({A}^*)\sim
\epsilon_-$ for the non-local charges. Using the convenient
parameterization of \cite{dualshG}, it is straightforward to
relate (\ref{K}) with the reflection matrix of \cite{GZ} using the
appropriate unitarization factor.\vspace{2mm}

$\bullet$ {\bf Massless limit: anisotropic Kondo and
Bazhanov-Lukyanov-Zamolodchikov models.}
The massless limit of the model (\ref{action}) can be reached in
various ways, according to the realization of the boundary
operators\,\footnote{I thank H. Saleur for stimulating
communications about this problem.} one would like to have. In the
dynamical boundary case, it should be reminded that the
integrability is preserved for boundary operators which satisfy
asymptotically (\ref{nonstand}) and the substitution
${A}\rightarrow \hE_+$, ${A}^*\rightarrow \hE_-$ with
$\rho=\rho^*=c^2(q^{1/2}-q^{-1/2})$ (see \cite{BK} for details).
The ``massless'' limit of (\ref{action}) is obtained if the
eigenvalues of $\mu^{1/2}\hE_\pm|_{\mu\rightarrow 0}$ are finite.
This can be done, for instance, by taking the limit
$v_0\rightarrow \infty$ in (\ref{Nomura2}) such that $v_0^2\mu
\sim 1$. It gives
\beqa A\rightarrow c_+ v_0 S_+ q^{s_3/2}\qquad \mbox{and}\qquad
A^*\rightarrow c_- v_0 S_- q^{-s_3/2}\label{kondolim} \ \eeqa
where (\ref{cpm}). Using (\ref{kondolim}) in (\ref{action}), the
resulting Hamiltonian becomes
\beqa H_{massless}\sim\frac{1}{8\pi}\int_{-\infty}^0 dx
\Big((\pi(x))^2 +(\partial_x\phi(x))^  
2\Big)-
\mu_B^{1/2}\big(S_+q^{s_3/2}e^{i\betah\phi(0)/2} +
S_-q^{-s_3/2}e^{-i\betah\phi(0)/2}\big) \ \label{kondo}\eeqa
For $q=1$, this Hamiltonian is the bozonized version of the
anisotropic Kondo model \cite{Kondo}. For $q\neq 1$, one obtains
the spin$-j$ generalization studied in details in
\cite{Fen96}.\vspace{1mm}

An other example of massless model is obtained using a different
realization of the Lax operator in (\ref{Kn}), and proceeding by
analogy to obtain (\ref{Nomura}). Taking the appropriate limit,
one obtains
\beqa A\rightarrow i(q^{1/2}-q^{-1/2})^{1/2}a_+ \qquad
\mbox{and}\qquad A^*\rightarrow
i(q^{1/2}-q^{-1/2})^{1/2}a_-q^{N/2}\ ,\label{osci}\eeqa
where the operators $a_+,a_-$ and $q^N$ generate the
$q-$oscillator algebra
\beqa a_-a_+ -q^{\pm1}a_+a_-=q^{\mp N} \qquad \mbox{and} \qquad
q^Na_{\pm}q^{-N} = q^{\pm 1}a_{\pm} \ . \eeqa
Notice that $q^{1/2}AA^* - q^{-1/2}A^*A=q^{1/2}-q^{-1/2}$. For
this realization of the boundary operators (in which case
 (\ref{qDG}) simplify to $q-$Serre relations), the model (\ref{action}) reduces to the one
proposed by Bazhanov-Lukyanov-Zamolodchikov in \cite{BLZ} at zero
voltage\,\footnote{In case of non-zero voltage, one can use a
slightly different expression for the generating function.}. This
model finds interesting application in non-equilibrium
systems.\vspace{2mm}

$\bullet$ {\bf Reflectionless points: Bassi-LeClair massive Kondo
model.}
The reflectionless points correspond to the values of the coupling
constant $\betah$ is such that $q=\pm 1$. In \cite{Bass99} a
rather general massive version of the anisotropic spin $1/2$ Kondo
model has been proposed. At the reflectionless points, it was
claimed to be integrable unlike the massless version which remains
integrable for arbitrary values of $\betah$. Using (\ref{Nomura}),
one can see that the Hamiltonian associated with (\ref{action})
coincides with the model proposed by Bassi and LeClair in
\cite{Bass99} for $q=1$. A more interesting property can
furthermore be exhibited: at this value of $q$, the boundary
operators $\hE_\pm$ and boundary non-local conserved charges
${\hat Q}^{(0)}_\pm$ generate the infinite dimensional Onsager
algebra (\ref{Onsager}). It is then possible to construct the
corresponding conserved quantities in involution which ensure the
integrability of the theory and confirms the conjecture of
\cite{Bass99}. Furthermore, this might open a way to extract exact
information (correlation functions,...) in the massive
regime.\vspace{2mm}

$\bullet$ {\bf Remark on the decoupling limit: Bulk/Boundary
``dual'' description.}
It is rather instructive to consider the limit in which the bulk
degrees of freedom (sine-Gordon field) and the boundary ones
$\hE_\pm(y)$ decouple. This can be realized either for Neumann
boundary conditions, or vanishing coupling constant $\betah=0$ in
(\ref{action}). For Neumann boundary conditions, i.e.
$\hE_\pm(y)=0$, only the bulk sine-Gordon contribution and the
term ${\cal A}_{boundary}$ survive. The non-local conserved
charges in this limit \cite{Mac} simplify to ${\hat
Q}^{(0)}_\pm|_{\hE_\pm\rightarrow Const.}$ in (\ref{nonloc}). On
the other hand, in the second case $\betah=0$, the action
(\ref{action}) becomes
\beqa {{\cal A}_{bsg}}|_{\betah=0}= \int_{-\infty}^{\infty}
dy\big({\cal E}_-(y) + {\cal E}_+(y)\big) + {\cal A}_{boundary}
\label{actionb}\ .\eeqa
In the asymptotic limit $(x,y)\rightarrow \pm \infty$, it is easy
to check that $\hE_\pm$ (for $\betah=0$) and ${\hat
Q}^{(0)}_\pm|_{\hE_\pm\rightarrow Const.}$ (for Neumann boundary
conditions) satisfy exactly the {\it same} algebraic relations
(\ref{qDG}). In view of the weak-strong coupling duality, in the
perturbative regime $\betah\rightarrow 0$ in (\ref{action}), one
can see that the boundary contribution (\ref{actionb}) actually
contains some informations about the nonperturbative regime which
is usually characterized by the non-local conserved charges. This
might explain why, for the analytic continuation
$\betah\rightarrow ib$ in (\ref{action}), the dynamical boundary
sinh-Gordon model is self-dual under weak-strong coupling duality
(the reflection amplitude of the fundamental particle is invariant
under the change $b\leftrightarrow 2/b$) \cite{BK}, contrary to
the model with fixed boundary conditions \cite{dualshG}.

\section{Concluding remarks}
The spectral parameter dependent reflection equation is usually
considered in the context of integrable systems with boundaries.
However, based on the analysis of this quadratic algebra we have
exhibited a new quantum integrable structure which is not specific
to boundary systems. The integrability condition was shown to be
associated with $q-$deformed Dolan-Grady relations (\ref{qDG}).
Our construction, which extends the one proposed by Dolan-Grady
some years ago \cite{DG}, leads to the existence of an (in)finite set
of conserved quantities in involution \cite{Tridiag}.In particular, it shows
that the concept of ``superintegrability'' can be further extended
for $q\neq 1$. In this direction, it is a challenge to find the
corresponding (in)finite dimensional $q-$Onsager algebra that
generate these conserved quantities in massive theories, and would
provide an alternative approach to massive quantum integrable
models. This new hidden symmetry (\ref{qDG}) was exhibited in
various bulk or boundary quantum integrable models  in which case
the operators $\textsf{A},\textsf{A}^*$ have been identified and
the first conserved quantities (\ref{qH}) and (\ref{G2})
constructed explicitly. Among the interesting applications, we
have shown that the sine-Gordon model enjoys a remarkable
dynamical symmetry associated with the quadratic Askey-Wilson
algebra. From this, we have obtained the asymptotic one-particle
states in terms of $q-$orthogonal polynomials, as well as the
structure of the boundary space of state in this model with a
boundary. From a more general point of view, we have also shown
that several examples of massive/massless boundary integrable
models are special cases of the model (\ref{action}). Finally, we
have pointed out a bulk-boundary ``dual'' relation in the
sine-Gordon model with a dynamical boundary. Although we didn't
discuss the XXZ spin chain with/without a boundary here, a similar
description also holds in this model.\vspace{1mm}

An interesting problem is wether the present approach can be
generalized to higher rank affine Lie algebras. Actually, there
has been some interest in generalizations of the Onsager algebra
and corresponding Dolan-Grady relations. Indeed, the Onsager
algebra is intimately related with $sl_2$. In case of $sl_n$, such
extension was studied in details in \cite{Uglov}. There, an
integrable Hamiltonian of the form
\beqa H=\kappa_1 \textsf{A}_1 + \kappa_2 \textsf{A}_2 + ... +
\kappa_n \textsf{A}_n\ ,\ \qquad n\geq 3 \eeqa
was proposed where $\kappa_i$ are arbitrary constants. This
Hamiltonian is a member of an infinite family of commuting
integrals of motion if the operators $\textsf{A}_i$ satisfy
certain trilinear relations. From our results, it is tempting to
consider a higher rank $q-$deformed generalization of (\ref{qDG})
with the following trilinear $q-$deformed relations:
\beqa [\textsf{A}_i,[\textsf{A}_i,\textsf{A}_j]_q]_{q^{-1}}=\rho_i
\textsf{A}_j \ ,\qquad \mbox{and}\qquad
[\textsf{A}_i,\textsf{A}_j]=0 \ \quad \mbox{if}\quad |i-j|>1\
\label{trili} \eeqa
where $\{i,j\}\in \{1,...,n\}$. In some sense, these relations
extend the concept of Leonard pair to any finite Lie algebra.
Obviously, using an appropriate shift $\textsf{A}_i\rightarrow
\textsf{A}_i + \alpha_i$ it is not difficult to generalize the
relations (\ref{nonstand}). For the same reasons as before, it is
consequently sufficient to focus on the relations (\ref{trili}).
For $q=1$ in (\ref{trili}), one obtains the relations found in
\cite{Uglov}. Also, for $\rho_i=0$, one recovers the $q-$Serre
relations associated with $U_q(sl_n)$. We intend to consider this
model together with applications to massive Toda field theories in
a separate publication.\vspace{2mm}

To conclude, we would like to stress that the power of the
relations (\ref{qDG}) lie in the algebraic statement which does
not refer to the number of dimensions or the the nature of the
space-time (continuous, discrete,...) ``hidden'' behind the
operators $\textsf{A},\textsf{A}^*$. An interesting problem is
consequently to find some examples where such hidden symmetry
appears explicitly in higher dimensions.\vspace{5mm}

{\underline{Note added:}}\\
Using the realization (\ref{Weyl}), it is easy to relate our
general expression (\ref{K}) with (\ref{solfin}) to known results.
For instance, choosing $c_0=\mp 1$, $c_1=\mp 2/(q-q^{-1})$,
$c_2=0$ and $q\rightarrow q^{-2}$ it is not difficult to check
that (\ref{solfin}) coincides with the one proposed in \cite{Kuzn}
(for $(-)$) (see also \cite{Delius}) or the one that follows from
\cite{BK} (for $(+)$).

Also, it should also be noted that the reflection matrix
(\ref{K}), for some special realizations of Leonard pairs in terms
of $U_{q^{1/2}}(\widehat{sl_2})$ can be related with a solution
proposed some years ago by A. Zabrodin in \cite{Zab}. This
solution was obtained from the non-dynamical $K-$matrix in
\cite{GZ} and $L$-operators satisfying (\ref{univYBA}). It should
be stressed that our solution (\ref{K}) is derived independently
of a realization of Leonard pairs, either directly from the RE
(\ref{RE}) or using the quantum affine reflection symmetry
described in Section 3.\vspace{3mm}

\vspace{0.5cm}

\noindent{\bf Acknowledgements:}  I am grateful to V. Bazhanov for
discussions at the preliminary stage of this work and P.
Terwilliger for explanations about his work on tridiagonal
algebras and Leonard pairs, as well as his advice in improving the
first version. I also thank S. Nicolis for explanations about the
Azbel-Hofstadter model, M. Niedermaier, P. Dorey and especially P.
Forgacs for discussions, interest in this work and important
comments about the manuscript. Part of this work is supported by
the TMR Network EUCLID ``Integrable models and applications: from
strings to condensed matter'', contract number
HPRN-CT-2002-00325.\vspace{1cm}

\centerline{\Large \bf Appendix}\vspace{0.5cm}

In this appendix, we want to show that an Hamitonian of the form
(\ref{qH}) can be obtained starting from an $L-$operator instead
of a $K-$operator. Indeed, in the quantum inverse scattering
method, an integrable lattice theory can be constructed
considering the monodromy matrix
\beqa T(u)=L_{{\cal V}_0\verb"N"}(u)...L_{{\cal
V}_0\verb"2"}(u)L_{{\cal V}_0\verb"1"}(u)\label{T} \eeqa
where $L_{{\cal V}_0\verb"j"}(u)$ acting on the quantum space
${\cal V}_j$ is a solution of the Yang-Baxter algebra (\ref{RLL})
for ${\cal V}_0={\cal V}'_0$ and ${\cal V}\rightarrow {\cal V}_j$.
Taking the trace over the auxiliary space ${\cal V}_0$ of
(\ref{T}), one obtains the transfer matrix
\beqa \tau(u)=Tr_{{\cal V}_0}(T(u))\label{gen}\eeqa
acting on $\otimes_{j=1}^{N} {\cal V}_j$. This generating function
provides a set of mutually commuting conserved quantities. For
instance, let us focus on the Yang-Baxter algebra (\ref{RLL}) with
(\ref{R}) associated with $U_{q^{1/2}}(\widehat{sl_2})$ in the
spin$-\frac{1}{2}$ auxiliary space representation ${\cal
V}_0={\cal V}'_0$ and $N=1$ in (\ref{T}). Following \cite{Wieg},
one can consider
\beqa L(u)=\left(
        \begin{array}{cc} -i\kappa^*{\rho^*}^{1/2}P -i\kappa \rho^{1/2}Q
        & ukq^{-1/4}(PQ^{-1})^{1/2}-u^{-1}k^{-1}q^{1/4}(PQ^{-1})^{-1/2}\\
   ukq^{1/4}(PQ^{-1})^{-1/2}-u^{-1}k^{-1}q^{-1/4}(PQ^{-1})^{1/2}  &
   -i\kappa{\rho}^{1/2}Q^{-1} -i \kappa^* {\rho^*}^{1/2} P^{-1} \\
\end{array}\right)\  \label{L} \eeqa
with parameters $\{k,\kappa,\kappa^*\}\in C\!\!\!\!I$. Using the
realization of the Leonard pair $\textsf{A},\textsf{A}^*$ in terms
of the Weyl algebra ${\cal W}_q$ (\ref{Weyl}), one obtains from
(\ref{gen}) a Hamiltonian acting on ${\cal V}$ of the form
\beqa H = (q-q^{-1})(\kappa \textsf{A} + \kappa^* \textsf{A}^*)\
\label{pH}\eeqa
where $\textsf{A},\textsf{A}^*$ satisfy the $q-$deformed
Dolan-Grady relations (\ref{qDG}).

\end{document}